\documentclass[useAMS,usenatbib,usegraphicx]{mn2e}
\usepackage[total={17.8cm,24.0cm},centering]{geometry}
\def\mgb{Mg$\,b$}
\def\hb{H$\beta$}
\def\galex{{\it GALEX}}
\def\sauron{{\tt SAURON}}
\title[The SAURON project - XVIII.]
 {The SAURON project - XVIII.\ The integrated UV--linestrength relations
 of early-type galaxies}
\author[M.\ Bureau et al.]
{Martin Bureau,$^{1}$\thanks{E-mail: bureau@astro.ox.ac.uk.}
 Hyunjin Jeong,$^{2}$ Sukyoung K.\ Yi,$^2$\thanks{E-mail:
   yi@yonsei.ac.kr} Kevin Schawinski,$^{3}$\thanks{Einstein Fellow.}
 \newauthor Ryan C.\ W.\ Houghton,$^1$ Roger L.\ Davies,$^1$ Roland
 Bacon,$^4$ Michele Cappellari,$^1$
 \newauthor P.\ Tim de Zeeuw,$^{5,6}$ Eric Emsellem,$^{4,5}$ Jes\'{u}s
 Falc\'{o}n-Barroso,$^7$ Davor Krajnovi\'{c},$^5$
 \newauthor Harald Kuntschner,$^8$ Richard M.\ McDermid,$^9$ Reynier
 F.\ Peletier,$^{10}$ Marc Sarzi,$^{11}$
 \newauthor Young-Jong Sohn,$^2$ Daniel Thomas,$^{12}$ Remco C.\ E.\
 van den Bosch,$^{13}$
 \newauthor and Glenn van de Ven$^{13}$\\
$^1$Sub-Department of Astrophysics, University of Oxford, Denys
 Wilkinson Building, Keble Road, Oxford OX1~3RH, U.K.\\
$^2$Department of Astronomy, Yonsei University, Seoul 120-749, Korea\\
$^3$Department of Physics, Yale University, New Haven, CT 06511,
 U.S.A.;\\
 Yale Center for Astronomy and Astrophysics, P.O.\ Box 208121,
 New Haven, CT 06520, U.S.A\\
$^4$Universit\'{e} de Lyon 1, CRAL, Observatoire de Lyon, 9 av.\ Charles Andr\'{e},
 F-69230 Saint-Genis Laval; CNRS, UMR 5574;\\
 ENS de Lyon, France\\
$^5$European Southern Observatory, Karl-Schwarzschild-Str.~2, 85748
 Garching, Germany\\
$^6$Leiden Observatory, Leiden University, Niels Bohrweg~2, 2333~CA Leiden,
 The Netherlands\\
$^7$Instituto de Astrof\'{i}sica de Canarias, V\'{i}a L\'{a}ctea s/n,
La Laguna, Tenerife, Spain;\\
Departamento de Astrof\'{i}sica, Universidad de La Laguna (ULL),
E-38205 La Laguna, Tenerife, Spain\\
$^8$Space Telescope European Coordinating Facility, European Southern
 Observatory, Karl-Schwarzschild-Str.~2, 85748 Garching, Germany\\
$^9$Gemini Observatory, 670 North A'Ohoku Place, Hilo, Hawaii 96720,
 U.S.A.\\
$^{10}$Kapteyn Astronomical Institute, University of Groningen, P.O.\ Box
 800, 9700 AV Groningen, The Netherlands\\
$^{11}$Centre for Astrophysics Research, University of Hertfordshire,
 Hatfield, U.K.\\
$^{12}$Institute of Cosmology and Gravitation, Mercantile House,
 Hampshire Terrace, University of Portsmouth, Portsmouth PO1 2EG, U.K.\\
$^{13}$Max Planck Institute for Astronomy, D-69117 Heidelberg, Germany}
\begin{document}
\maketitle
%
%
\begin{abstract} Using far (FUV) and near (NUV) ultraviolet photometry
  from guest investigator programmes on the {\it Galaxy Evolution
    Explorer} (\galex) satellite, optical photometry from the MDM
  Observatory and optical integral-field spectroscopy from \sauron, we
  explore the UV--linestrength relations of the $48$ nearby early-type
  galaxies in the \sauron\ sample. Identical apertures are used for
  all quantities, avoiding aperture mismatch. We show that galaxies
  with purely old stellar populations show well-defined correlations
  of the integrated FUV$-V$ and FUV$-$NUV colours with the integrated
  \mgb\ and \hb\ absorption linestrength indices, strongest for
  FUV$-$NUV. Correlations with the NUV$-V$ colour, Fe5015 index and
  stellar velocity dispersion $\sigma$ are much weaker. These
  correlations put stringent constraints on the origin of the
  UV-upturn phenomenon in early-type galaxies, and highlight its
  dependence on age and metallicity. In particular, despite recent
  debate, we recover the negative correlation between FUV$-V$ colour
  and Mg linestrength originally publicised by \citet*{bbbfl88}, which
  we refer to as the ``Burstein relation'', suggesting a positive
  dependence of the UV-upturn on metallicity. We argue that the
  scatter in the correlations is real, and present mild evidence that
  a strong UV excess is preferentially present in slow-rotating
  galaxies. We also demonstrate that most outliers in the correlations
  are galaxies with current or recent star formation, some at very low
  levels. We believe that this sensitivity to weak star formation,
  afforded by the deep and varied data available for the \sauron\
  sample, explains why our results are occasionally at odds with other
  recent but shallower surveys. This is supported by the analysis of a
  large, carefully-crafted sample of more distant early-type galaxies
  from the Sloan Digital Sky Survey (SDSS), more easily comparable
  with current and future large surveys.
\end{abstract}
%
%
\nobreak
\section{INTRODUCTION}
\label{sec:intro}
Far-ultraviolet (FUV) radiation was first discovered in early-type
galaxies by the {\it Orbiting Astronomical Observatory-2} in 1969
\citep{cwp72}. It was rapidly realised that, below $\approx2000$~\AA,
many early-types show spatially extended {\em rising} flux for
decreasing wavelengths \citep[e.g.][]{cw79,bcho80}, a surprising
behaviour for presumably old stellar populations. This spawned a
debate on the origin of this UV upturn that continues to this
day.

The leading hypothesis suggests that this flux is due to hot stars on
an extreme horizontal branch (EHB) and their progeny (e.g.\ asymptotic
giant branch (AGB) stars manqu\'{e}), either metal-poor (e.g.,
\citealt*{ldz94,pl97}; see also \citealt{bg08}) or metal-rich
\citep*[e.g.][]{gr90,hdp92,bcf94,dor95,ydk97}, with possible
significant effects from He enhancement
\citep[e.g.][]{letal05b,ksoyly07}. Binary star systems offer another
compelling possibility, with limited dependence on age and metallicity
\citep*[e.g.][]{mhmn01,hpmm03,hpl07}. Post-AGB (PAGB) stars in the
planetary nebula stage also emit FUV radiation, but they are generally
thought to be too hot, too few and to have too short a lifetime to
explain the observed fluxes alone
\citep[e.g.][]{gr90,dor95,bfdd97}. Clearly, very young stars in
star-forming galaxies can also emit in the FUV \citep[see,
e.g.,][]{fs00,yetal05}, but they necessarily have evolving properties
(e.g.\ characteristic temperature of the UV spectrum), they have
different spectral signatures from those normally observed in
UV-upturn galaxies (e.g.\ strong C~{\small IV} lines) and they are
expected to be clumpy, none of which is supported by observations. A
thorough review of these issues is available in \citet{o99}, while
\citet{b04} and \citet{y08} provide more recent but shorter
overviews. See \citet{h09} for a review of hot subdwarf stars.

In addition to its roughly constant spectral slope ($T_{\rm
  eff}\approx20,000$~K; \citealt{bfdd97}), one of the main constraints
on the possible origins of the UV upturn is its systematic dependence
on bulk galaxy properties. Although mentioned in \citet{f83},
\citeauthor*{bbbfl88} (\citeyear{bbbfl88}, hereafter
\citeauthor{bbbfl88}) are generally considered the first to have
systematically investigated the dependence of the UV upturn on the
stellar population and dynamical properties of a large sample of
early-type galaxies. \citeauthor{bbbfl88} found clear non-linear
correlations of the UV-optical colours with both the absorption
linestrength index Mg$_{2}$ \citep[see][]{ffbg85} and the central
stellar velocity dispersion $\sigma$. In memory of David Burstein and
in view of the long-lasting impact of \citeauthor{bbbfl88}, we will
hereafter refer to the former correlation as the ``Burstein
relation''. These correlations have formed the basis of much of the
observational and theoretical work on the UV upturn to
date. Crucially, contrary to broadband optical and near-infrared
colours, the Mg$_{2}$ linestrength was found to be stronger for {\em
  bluer} UV-optical colours, i.e.\ galaxies become bluer with
increasing metallicity in the UV. Globular clusters show different
correlations, pointing to a slightly different origin for the UV
radiation (e.g.\ \citealt{dor95,soklbbfr06}; but see also
\citealt*{lgc07}).

The recent availability of large-area, deep UV imaging from the {\it
  Galaxy Evolution Explorer} (\galex) satellite has added another
layer of complexity to the problem. Observations of large samples of
early-type galaxies from the Sloan Digital Sky Survey (SDSS) have
revealed that a very large fraction of them (over $30\%$) have
remarkably strong near-ultraviolet (NUV) and FUV emission.
\citet{yetal05} initially concluded that the majority of UV-blue
early-type galaxies were inconsistent with exhibiting a classic UV
upturn, and more likely experienced residual star formation.
\citet{setal07} and \citet{ketal07} elaborated on this with larger
samples, the main conclusion being that residual star formation of a
few percent of the stellar mass is very common among early-type
galaxies. The UV emission regularly found in early-type galaxies is
simply too powerful to be accounted for by any current model of the UV
upturn, and is far more powerful than anything found in traditional
UV-upturn galaxies such as NGC~1399 and NGC~4552.

Some recent work has also thrown the results of \citeauthor{bbbfl88}
into doubt. In particular, for a very large sample of local early-type
galaxies with \galex\ UV imaging and SDSS optical spectroscopy,
\citet{retal05} report no correlation of the FUV$-r$ colour against
either the metallicity-sensitive Mg$_2$ and D4000 indices or the
stellar velocity dispersion, even after attempting to remove
non-early-type and star-forming galaxies from their sample. If
anything, a slight correlation opposite to the Burstein relation is
observed. \citet*{dbd02} report similar findings, although without
active galactic nucleus (AGN) or star-formation cuts and using a UV
filter centred on $2000$~\AA, where the UV upturn phenomenon may not
dominate. Using data on nearby galaxies but sorting according to
morphological type, \citet{detal07} do find correlations for
elliptical galaxies but those for lenticulars are weak at best, with
much scatter and many outliers. \citet{betal05} recover weak
correlations analogous to those of \citeauthor{bbbfl88} for Virgo
Cluster early-type members, but it is unclear whether the correlations
extend to low-luminosity systems.
If these recent results prove to be robust, much of the theoretical
work on the UV upturn will need to be revised, particularly its
dependence on metallicity.

To sort out possible discrepancies between the above, as much
information as possible is required on the sample galaxies under
consideration, particularly regarding possible low-level star
formation. We thus consider two samples here. First, the \sauron\
sample of $48$ nearby elliptical and lenticular galaxies (see
Table~\ref{tab:sauron}; \citealt*{zetal02}, hereafter
\citeauthor{zetal02}), for which detailed integral-field maps of the
stellar kinematics (\citealt*{eetal04}, hereafter
\citeauthor{eetal04}), stellar absorption linestrengths and
populations (\citealt*{ketal06}, hereafter \citeauthor{ketal06};
\citealt*{ketal10}, hereafter \citeauthor{ketal10}) and ionised gas
distribution and kinematics (\citealt*{setal06}, hereafter
\citeauthor{setal06}; \citealt*{setal10}, hereafter
\citeauthor{setal10}) are available, along with detailed dynamical
modeling (e.g.\ \citealt*{cetal06,cetal07,eetal07}, hereafter
\citeauthor{cetal06}, \citeauthor{cetal07} and \citeauthor{eetal07},
respectively). The molecular gas content \citep*{cyb07,cbyc10}, UV
emission (\citealt*{jetal09}, hereafter \citeauthor{jetal09}) and
mid-infrared (MIR) emission (\citealt*{tbm09}; \citealt*{shetal10},
hereafter \citeauthor{shetal10}) of the \sauron\ sample have also been
investigated, to more directly probe its star formation
properties. Second, we construct a SDSS sample similar to those of
\citet{retal05} and \citet{detal07}, but apply much more stringent
cuts to remove star-forming galaxies and low quality data. In both
cases, we recover the relationships originally suggested by
\citeauthor{bbbfl88} for the non-star-forming (i.e.\ quiescent)
galaxies only, while including galaxies with even a very low level of
star formation considerably blurs the relations or can even give rise
to trends in the opposite sense.

In Section~\ref{sec:data}, we present the photometric and
spectroscopic data and related analysis. Section~\ref{sec:corrs}
presents the UV--linestrength relations for the \sauron\ sample while
Section~\ref{sec:sf} discusses the effects of star formation on
them. The SDSS sample is presented and discussed in
Section~\ref{sec:sdss}. We bring all our results together in
Section~\ref{sec:discussion} and conclude briefly in
Section~\ref{sec:conclusions}.
%
%
\section{OBSERVATIONS AND DATA REDUCTION}
\label{sec:data}
Although spatially-resolved data provide in principle a more powerful
discriminant of UV-upturn theories, we focus here on integrated
quantities as they provide a link with past work (particularly
\citeauthor{bbbfl88}) and offer a crucial benchmark for future studies
at high redshifts. The latter in particular are essential to probe the
evolution of the UV-upturn with look-back time (and thus turnoff
mass), an important test of single versus binary stellar origins
\citep[see, e.g.,][]{ylwpdo99,letal05a,retal07}. The two-dimensional
nature of the \sauron\ data will be more fully exploited in a companion
paper (Jeong et al., in preparation).

\citeauthor{ketal06} showed that, to first order, iso-index contours
follow a galaxy light (although they are slightly flatter than the
isophotes for \mgb\ in $\approx40\%$ of the cases). We therefore adopt
elliptical apertures to measure integrated quantities throughout this
paper. We will generally refer to quantities (magnitudes and
linestrengths) integrated within ellipses of semi-major axis $R_{\rm
  e}/2$, where $R_{\rm e}$ is the effective radius of the galaxy
(i.e.\ the radius of the aperture encompassing half the galaxy
light). A radius of $R_{\rm e}/2$ is ideal for our work as it is large
enough to accurately represent global values, yet it is small enough
that the corresponding elliptical apertures are fully covered by the
\sauron\ field-of-view (FOV) for most objects ($\approx80\%$). Other
interesting radii include $R_{\rm e}/8$ (i.e.\ central values) and
$R_{\rm e}$ (global values), although the latter requires a small
extrapolation for $\approx60\%$ of the objects.
%
%

\begin{table*}
\caption{\sauron\ sample data}
\label{tab:sauron}
\begin{tabular}{@{}lrrrrrrrrr}
\hline
Galaxy & FUV$_{R{\rm e}/2}$ & NUV$_{R{\rm e}/2}$ & $V_{R{\rm e}/2}$ &
$E(B-V)$ & \mgb$_{R{\rm e}/2}$ & Fe5015$_{R{\rm e}/2}$ & \hb$_{R{\rm e}/2}$ &
$\sigma_{\rm e}$ & Survey \\
 & (mag) & (mag) & (mag) & (mag) & (\AA) & (\AA) & (\AA) & (km~s$^{-1}$) \\
(1) & (2) & (3) & (4) & (5) & (6) & (7) & (8) & (9) & (10) \\\hline
NGC0474  &  19.46 $\pm$ 0.18  &  17.92 $\pm$ 0.07  &  12.70 $\pm$ 0.05  &  0.034  &  3.65  &  4.65  &  1.87  &  142  &   N  \\
NGC0524  &  18.17 $\pm$ 0.21  &  16.99 $\pm$ 0.12  &  11.21 $\pm$ 0.03  &  0.083  &  4.22  &  5.07  &  1.60  &  225  &   G  \\
NGC0821  &  18.96 $\pm$ 0.26  &  17.40 $\pm$ 0.10  &  11.88 $\pm$ 0.03  &  0.110  &  3.90  &  4.54  &  1.71  &  182  &   G  \\
NGC1023  &  17.45 $\pm$ 0.07  &  16.19 $\pm$ 0.04  &  10.61 $\pm$ 0.02  &  0.061  &  4.26  &  4.85  &  1.62  &  165  &   N  \\
NGC2695  &  19.17 $\pm$ 0.08  &  18.28 $\pm$ 0.06  &  12.87 $\pm$ 0.04  &  0.018  &  4.08  &  4.17  &  1.41  &  184  &   G  \\
NGC2699  &  19.98 $\pm$ 0.10  &  18.91 $\pm$ 0.05  &  13.31 $\pm$ 0.05  &  0.020  &  3.75  &  4.74  &  1.87  &  123  &   G  \\
NGC2768  &  18.15 $\pm$ 0.16  &  16.82 $\pm$ 0.08  &  11.25 $\pm$ 0.04  &  0.044  &  3.81  &  4.40  &  1.70  &  200  &   N  \\
NGC2974  &  18.54 $\pm$ 0.07  &  17.64 $\pm$ 0.05  &  12.10 $\pm$ 0.04  &  0.054  &  4.23  &  4.84  &  1.77  &  227  &   G  \\
NGC3032  &  16.53 $\pm$ 0.01  &  15.79 $\pm$ 0.02  &  13.25 $\pm$ 0.05  &  0.017  &  1.79  &  3.60  &  4.46  &   90  &   G  \\
NGC3156  &  18.70 $\pm$ 0.22  &  17.36 $\pm$ 0.08  &  13.35 $\pm$ 0.03  &  0.034  &  1.79  &  3.47  &  3.39  &   66  &   A  \\
NGC3414  &  18.01 $\pm$ 0.24  &  17.02 $\pm$ 0.10  &  12.03 $\pm$ 0.03  &  0.024  &  3.97  &  4.43  &  1.61  &  191  &   A  \\
NGC3489  &  18.24 $\pm$ 0.10  &  16.56 $\pm$ 0.03  &  11.81 $\pm$ 0.02  &  0.017  &  2.49  &  4.32  &  2.74  &   99  &   A  \\
NGC3608  &  18.26 $\pm$ 0.40  &  17.10 $\pm$ 0.16  &  12.02 $\pm$ 0.04  &  0.021  &  3.90  &  4.62  &  1.73  &  167  &   A  \\
NGC4150  &  18.73 $\pm$ 0.04  &  17.46 $\pm$ 0.02  &  13.01 $\pm$ 0.03  &  0.018  &  2.26  &  3.81  &  3.22  &   77  &   N  \\
NGC4262  &  19.50 $\pm$ 0.08  &  18.74 $\pm$ 0.04  &  13.37 $\pm$ 0.02  &  0.035  &  4.15  &  4.40  &  1.54  &  164  &   A  \\
NGC4278  &  17.01 $\pm$ 0.03  &  16.40 $\pm$ 0.03  &  11.29 $\pm$ 0.02  &  0.029  &  4.42  &  4.34  &  1.56  &  217  &   N  \\
NGC4374  &  16.52 $\pm$ 0.07  &  15.54 $\pm$ 0.05  &  10.19 $\pm$ 0.03  &  0.040  &  4.27  &  4.54  &  1.52  &  261  &   N  \\
NGC4387  &  20.20 $\pm$ 0.14  &  18.66 $\pm$ 0.06  &  13.22 $\pm$ 0.02  &  0.033  &  3.72  &  4.40  &  1.69  &   98  &   N  \\
NGC4458  &  20.21 $\pm$ 0.26  &  18.40 $\pm$ 0.08  &  13.07 $\pm$ 0.05  &  0.024  &  3.36  &  3.84  &  1.71  &   83  &   N  \\
NGC4459  &  17.64 $\pm$ 0.06  &  16.41 $\pm$ 0.04  &  11.41 $\pm$ 0.03  &  0.046  &  3.60  &  4.54  &  2.02  &  155  &   G  \\
NGC4473  &  18.35 $\pm$ 0.06  &  17.09 $\pm$ 0.03  &  11.45 $\pm$ 0.02  &  0.028  &  4.18  &  4.76  &  1.62  &  186  &   N  \\
NGC4477  &  18.09 $\pm$ 0.11  &  16.87 $\pm$ 0.06  &  11.46 $\pm$ 0.04  &  0.032  &  4.00  &  4.69  &  1.67  &  147  &   N  \\
NGC4486  &  15.12 $\pm$ 0.06  &  14.59 $\pm$ 0.07  &   9.76 $\pm$ 0.04  &  0.022  &  4.74  &  4.38  &  1.27  &  268  &   N  \\
NGC4526  &  17.59 $\pm$ 0.05  &  16.12 $\pm$ 0.03  &  11.04 $\pm$ 0.03  &  0.022  &  4.23  &  4.82  &  1.77  &  214  &   G  \\
NGC4546  &  18.87 $\pm$ 0.07  &  17.41 $\pm$ 0.04  &  11.70 $\pm$ 0.03  &  0.034  &  4.18  &  4.60  &  1.63  &  189  &   N  \\
NGC4550  &  19.08 $\pm$ 0.05  &  18.02 $\pm$ 0.03  &  13.18 $\pm$ 0.02  &  0.039  &  3.03  &  4.18  &  2.14  &  103  &   N  \\
NGC4552  &  16.45 $\pm$ 0.02  &  16.03 $\pm$ 0.03  &  11.05 $\pm$ 0.02  &  0.041  &  4.60  &  5.07  &  1.58  &  233  &   N  \\
NGC4564  &  18.15 $\pm$ 0.04  &  17.45 $\pm$ 0.03  &  12.27 $\pm$ 0.04  &  0.035  &  4.24  &  4.92  &  1.67  &  150  &   N  \\
NGC4570  &  18.85 $\pm$ 0.04  &  17.67 $\pm$ 0.03  &  12.18 $\pm$ 0.03  &  0.022  &  4.11  &  4.67  &  1.56  &  167  &   G  \\
NGC4621  &  17.26 $\pm$ 0.05  &  16.27 $\pm$ 0.04  &  10.94 $\pm$ 0.04  &  0.033  &  4.30  &  4.64  &  1.56  &  200  &   G  \\
NGC4660  &  19.21 $\pm$ 0.23  &  17.90 $\pm$ 0.08  &  12.41 $\pm$ 0.02  &  0.033  &  4.24  &  4.70  &  1.59  &  181  &   A  \\
NGC5198  &  19.31 $\pm$ 0.10  &  18.42 $\pm$ 0.07  &  13.06 $\pm$ 0.03  &  0.023  &  4.13  &  4.65  &  1.64  &  173  &   G  \\
NGC5308  &  20.41 $\pm$ 0.08  &  19.21 $\pm$ 0.04  &  13.49 $\pm$ 0.02  &  0.018  &  4.34  &  4.78  &  1.58  &  201  &   G  \\
NGC5813  &  18.25 $\pm$ 0.16  &  17.07 $\pm$ 0.09  &  11.87 $\pm$ 0.05  &  0.057  &  4.26  &  4.66  &  1.64  &  210  &   G  \\
NGC5831  &  19.09 $\pm$ 0.18  &  17.64 $\pm$ 0.09  &  12.30 $\pm$ 0.04  &  0.059  &  3.59  &  4.66  &  1.89  &  148  &   G  \\
NGC5838  &  18.47 $\pm$ 0.08  &  17.30 $\pm$ 0.04  &  12.05 $\pm$ 0.03  &  0.053  &  4.25  &  4.85  &  1.68  &  232  &   G  \\
NGC5845  &  19.99 $\pm$ 0.08  &  18.86 $\pm$ 0.04  &  13.29 $\pm$ 0.05  &  0.053  &  4.36  &  4.96  &  1.65  &  237  &   N  \\
NGC5846  &  17.02 $\pm$ 0.14  &  16.20 $\pm$ 0.10  &  11.11 $\pm$ 0.05  &  0.055  &  4.54  &  4.78  &  1.42  &  213  &   N  \\
NGC5982  &  18.55 $\pm$ 0.06  &  17.66 $\pm$ 0.04  &  12.31 $\pm$ 0.03  &  0.018  &  4.13  &  5.05  &  1.74  &  223  &   G  \\
NGC7332  &  21.91 $\pm$ 0.79  &  18.70 $\pm$ 0.10  &  12.78 $\pm$ 0.01  &  0.037  &  3.33  &  4.97  &  2.29  &  125  &   A  \\
NGC7457  &  19.32 $\pm$ 0.42  &  16.99 $\pm$ 0.10  &  11.98 $\pm$ 0.04  &  0.052  &  2.83  &  4.36  &  2.36  &   75  &   G  \\
\hline
\end{tabular}

Columns: (2)--(4) and (6)--(8): Quantities integrated within
elliptical apertures of semi-major radii $R_{\rm e}/2$; (2)--(4) have
been corrected for Galactic extinction using (5) and the errors are
based on Poisson statistics and our sky subtraction. The adopted formal random
error on the values in (6)--(8) is $0.1$~\AA\ (see \citeauthor{ketal06}). (9) Effective
stellar velocity dispersion within $R_{\rm e}$ from
\citeauthor{ketal10}, with an adopted error of $5\%$. (10)
Origin of data: A=All-sky Imaging Survey, G=own guest investigator
programmes, N=Nearby Galaxy Survey.
\end{table*}
%
%
\subsection{UV photometry}
\label{sec:data_uv}
All $48$ \sauron\ early-type galaxies except seven were observed in
FUV ($1350$--$1750$~\AA) and/or NUV ($1750$--$2750$~\AA) with the
Medium imaging mode of \galex\ \citep[see][]{maetal05,moetal05}. A
total of $17$ galaxies were observed as part of our own UV imaging
survey of the \sauron\ sample (\galex\ guest investigator programmes
GI1\_109 and GI3\_041) and another $17$ as part of the \galex\ team
Nearby Galaxy Survey (NGS; \citealt{getal05}). All these objects were
observed for at least one orbit (typically $1500$--$1700$~s) and have
excellent data for any aperture up to $R_{\rm e}$. The remaining $7$
galaxies have only short exposures (typically $100$--$150$~s) from the
\galex\ team All-sky Imaging Survey (AIS; see \citealt{maetal05}), and
magnitudes integrated within $R_{\rm e}$ are less reliable. The
resulting sample is listed in Table~\ref{tab:sauron}. The data
reduction, analysis and products are described more fully in a
separate paper (\citeauthor{jetal09}), so we only summarise our
processing here.

We used the pre-processed images delivered from the \galex\ pipeline
but performed our own sky subtraction by measuring the background level
in source-free regions of the images. The spatial resolution is
roughly $4\farcs5$ and $6\farcs0$ FWHM at FUV and NUV, respectively,
sampled with $1\farcs5\times1\farcs5$ pixels. To avoid spurious colour
gradients (especially in the inner parts), we thus convolved the FUV
data to the spatial resolution of the NUV observations before any
analysis.

We carried out surface photometry by measuring the surface brightness
within elliptical annuli in the standard manner, using the {\small
  ELLIPSE} task in {\small IRAF} (Image Reduction and Analysis
Facility). To derive accurate colours, the ellipses were fitted to the
NUV images only (with the better signal-to-noise ratio S/N), and those
ellipses were then imposed on the FUV images. Photometric zero-points
were taken from \citet{moetal05} and the data were corrected for
Galactic extinction using the extinction law of \citet*{ccm89}. We
take the $E(B-V)$ value for each galaxy as tabulated in the NASA/IPAC
Extragalactic Database (NED) from \citet*{sfd98}. For this paper, only
the FUV and NUV surface brightness profiles integrated within $R_{\rm
  e}/2$ are discussed, but we have checked that analogous results
stand with apertures of $R_{\rm e}/8$ and $R_{\rm e}$. The integrated
UV magnitudes corrected for Galactic extinction are listed in
Table~\ref{tab:sauron}, along with errors based on Poisson statistics
and our sky subtraction (all UV magnitudes are AB magnitudes).
%
%
\subsection{Optical photometry}
\label{sec:data_mdm}
Optical images in the {\it Hubble Space Telescope} ({\it HST}) filter
F555W (similar to Johnson $V$) were obtained for the entire \sauron\
sample using the MDM Observatory 1.3-m McGraw-Hill Telescope over
several observing runs in 2002--2005. The field-of-view of the MDM
images is $17\farcm3\times17\farcm3$ with $0\farcs508\times0\farcs508$
pixels, permitting excellent sky subtraction and proper sampling of
the seeing. The MDM observations are described in Fal\'{c}on-Barroso
et al.\ (in preparation). They were reduced in the standard manner in
{\small IRAF}. The photometric calibration was obtained by observing
photometric standard stars from \citet{l92} over a range of air
masses, resulting in a systematic calibration uncertainty of
$0.03$~mag. Like the FUV data, the MDM images were convolved to the
spatial resolution of the NUV data before any analysis.

Again like the FUV data, the surface brightness profiles for the MDM
data were measured only after imposing the NUV ellipses, ensuring
accurate UV--optical colours. Magnitudes integrated within elliptical
apertures of semi-major axis $R_{\rm e}/2$ (identical to the UV
apertures) were then calculated and corrected for Galactic
extinction. The integrated and corrected $V$ magnitudes (and errors)
are listed in Table~\ref{tab:sauron}.
%
%
\subsection{\sauron\ linestrengths and kinematics}
\label{sec:data_sauron}
All linestrength measurements come from the \sauron\ data published in
\citeauthor{ketal06}, updated in \citeauthor{ketal10}. The
observations and data reduction were thoroughly discussed there and
will not be repeated here. While \citeauthor{bbbfl88} strived to
derive integrated $V$ magnitudes within apertures equivalent to the
roughly $11\arcsec\times22\arcsec$ rectangular aperture of the {\it
  International Ultraviolet Observer} ({\it IUE}), they had no choice
but to use essentially central measurements from slit spectroscopy for
Mg$_2$ and $\sigma$. A significant improvement in the current work is
that integral-field spectroscopy allows us to measure linestrength
indices over exactly the same apertures as the UV and optical
photometry.

\citeauthor{ketal10} lists linestrength values integrated within
circular apertures of radii $R_{\rm e}/8$ and $R_{\rm e}$, but we
again use here NUV-defined elliptical apertures of semi-major axes
$R_{\rm e}/2$, as done in Sections~\ref{sec:data_uv} and
\ref{sec:data_mdm}, to avoid potential aperture mismatch problems. We
use simple luminosity-weighted linestrength averages of all the bins
within the apertures. The resulting integrated linestrength values are
listed in Table~\ref{tab:sauron}. While not formally equivalent, we
show in Appendix~\ref{app:averages} that these measurements are as
robust as those obtained by summing the spectra of all the bins
concerned and rederiving the linestrengths from scratch. We therefore
adopt the same uncertainty for our integrated linestrengths as
\citeauthor{ketal10}, a formal random error of $0.1$~\AA\ (maximum)
for all measurements.

Since linestrengths are known to correlate with the central velocity
dispersion of galaxies \citep[e.g.][]{tdfb81,bddfl88,betal03},
following \citeauthor{bbbfl88} we also explore the dependence of the
UV--optical colours on $\sigma$. For that, we do not use central
velocity dispersion measurements as is usually done, but rather again
use properly integrated measurements obtained by summing all the
spectra within a given aperture, and re-measuring the stellar
kinematics (assuming a pure Gaussian line-of-sight velocity
distribution (LOSVD), i.e.\ no high order Gauss-Hermite terms). Such
stellar velocity dispersion measurements are already available in
\citeauthor{cetal06} (updated in \citeauthor{ketal10}) for a circular
aperture of radius $R_{\rm e}$, and the extrapolation required when
the field-of-view of \sauron\ does not fully cover $R_{\rm e}$ is
detailed there. The resulting ``effective'' stellar velocity
dispersion $\sigma_{\rm e}$ is truly the luminosity-weighted second
moment of the LOSVD within $R_{\rm e}$, and is a good approximation to
the second velocity moment which appears in the virial
equation. Integrated velocity dispersions only weakly depend on the
aperture used, so we do not re-calculate new values for the elliptical
apertures used throughout this paper, but instead simply adopt the
values and uncertainties listed in \citeauthor{ketal10} (see
Table~\ref{tab:sauron}).
%
%
\section{UV RELATIONS}
\label{sec:corrs}
When exploring possible correlations between stellar absorption
linestrengths and broadband photometry, our main advantage over all
previous studies, including \citeauthor{bbbfl88}, \citet{retal05} and
\citet{detal07}, is the availability not only of UV imaging, but most
importantly of integral-field spectroscopy for all targets. All
photometric and spectroscopic measurements are thus made on identical
apertures, as described in Section~\ref{sec:data}.

Figure~\ref{fig:fuv-v} shows the distance-independent correlations
between the FUV$-V$ colour and the linestrength indices \mgb, Fe5015
and \hb. Although the transformation of \mgb, Fe5015 and \hb\ to
stellar age, metallicity and $\alpha$-element (over-)abundance is
model dependent, \hb\ is primarily a tracer of age. For the range of
metallicities present in early-type galaxies and considering various
stellar population synthesis models \citep[e.g.][]{tmb03,s07}, an \hb\
linestrength of $1.8$~\AA\ corresponds to single stellar population
luminosity-weighted ages of roughly $5$ to $10$~Gyrs (see, e.g.,
Figure~3 in \citeauthor{ketal10}). This linestrength value thus
represents a good choice to separate young from old galaxies, or
rather primarily old galaxies with a significant amount of recent star
formation and young stars from primarily old galaxies with no or a
minimal additional young stellar population. Changing this threshold
value slightly does not significantly affect any of our results. To
highlight the effects of recent star formation in
Figure~\ref{fig:fuv-v}, galaxies with \hb$\,>1.8$~\AA, that are
inconsistent with a purely old stellar population, are therefore
plotted as crosses. Galaxies with no or minimal (in a
luminosity-weighted sense) young stellar populations
(\hb$\,\le1.8$~\AA) are plotted as filled circles.

There is no physical reason why a linear relation should exist between
linestrengths and UV excess, but linear fits to the data with
\hb$\,\le1.8$~\AA\ are shown as solid lines to guide the eye. The
value of the correlation coefficient is reported in the bottom right
corner of each panel, and the parameters of the best-fit lines are
reported in Table~\ref{tab:linear_fits}, all taking errors properly
into account. Lastly, the symbols are colour-coded according to the
effective stellar velocity dispersion $\sigma_{\rm e}$, increasing
from purple to red (\citeauthor{ketal10}).
%
%

\begin{table}
\caption{Parameters of the best-fit linear UV--linestrength relations
  for old galaxies only}
\label{tab:linear_fits}
\begin{tabular}{@{}llrrr}
\hline
Colour    & Line  & Slope            & Zero-point   & Scatter \\
(mag)     & (\AA) & (\AA\ mag$^{-1}$) & (\AA)        & (\AA)   \\\hline
FUV$-V$   & \mgb  & $-0.40\pm0.09$   & $6.75\pm0.57$ & $0.15$ \\
          & Fe    & $-0.07\pm0.11$   & $5.11\pm0.75$ & $0.22$ \\
          & \hb   & $ 0.12\pm0.05$   & $0.83\pm0.32$ & $0.08$ \\
FUV$-$NUV & \mgb  & $-0.64\pm0.12$   & $4.89\pm0.14$ & $0.15$ \\
          & Fe    & $-0.21\pm0.17$   & $4.89\pm0.19$ & $0.21$ \\
          & \hb   & $ 0.20\pm0.08$   & $1.40\pm0.09$ & $0.07$ \\
\hline
\end{tabular}
\end{table}
%
%
\subsection{Burstein relation: FUV$-V$ vs.\ \mgb}
\label{sec:corrs_fuv-v-mgb}
\begin{figure}
\begin{center}
\includegraphics[width=7.45cm,clip=]{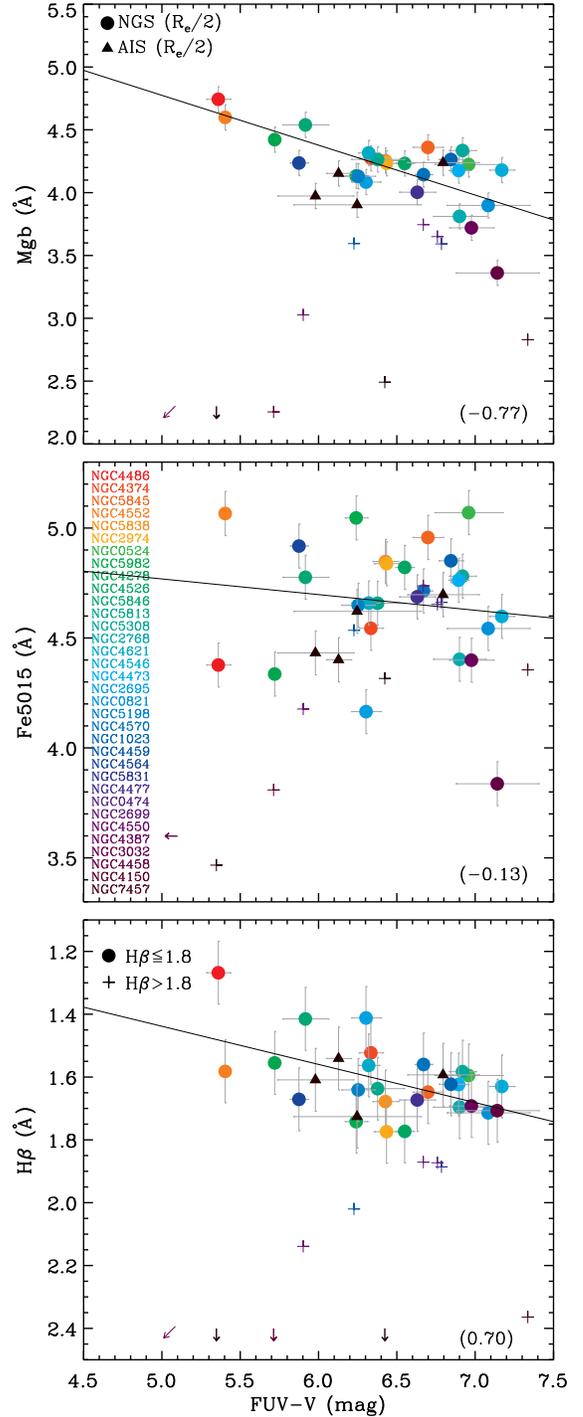}
\caption{UV upturn dependence on stellar absorption linestrengths. The
  \mgb\ ({\em top}), Fe5015 ({\em middle}) and \hb\ ({\em bottom})
  linestrengths are shown as a function of the FUV$-V$ colour, all
  integrated within $R_{\rm e}/2$. The top plot is our own version of
  the Burstein relation. Objects inconsistent with a purely old
  stellar population (\hb$>1.8$~\AA) are plotted as crosses, others as
  filled circles. The location of data points lying outside the figure
  boundaries is indicated by arrows. Note the inverted \hb\
  axis. Linear fits to the data with \hb$\,\le1.8$~\AA\ are shown as
  solid lines to guide the eye, and the correlation coefficient is
  reported in the bottom right corner of each panel. The best-fit
  parameters are listed in Table~\ref{tab:linear_fits}. The colour
  table traces the effective stellar velocity dispersion $\sigma_{\rm
    e}$, from purple at small dispersions through green and red at
  high dispersions.}
\label{fig:fuv-v}
\end{center}
\end{figure}

The top panel of Figure~\ref{fig:fuv-v} shows the (FUV$-V$)--\mgb\
relation, analogous to the (1550$-V$)--Mg$_2$ relation discussed by
\citeauthor{bbbfl88}, although both the FUV filter and the Mg index
are slightly different and we use a common aperture of $R_{\rm
  e}/2$. This is thus our own version of the Burstein
relation. Reassuringly, we see a very similar trend. Most importantly,
there is a clear correlation between the FUV$-V$ colour and the \mgb\
linestrength for galaxies with no significant young stellar population
(\hb$\,\le1.8$~\AA, i.e.\ quiescent galaxies). This is the systematic
variation of the hot component of old stellar populations discovered
by \citeauthor{bbbfl88}. Galaxies with a substantial young stellar
population (\hb$\,>1.8$~\AA) blur the relation and are found at
systematically lower \mgb\ values, although over a wide range of
FUV$-V$ colour.
%
%

\begin{figure}
\begin{center}
\includegraphics[width=7.45cm,clip=]{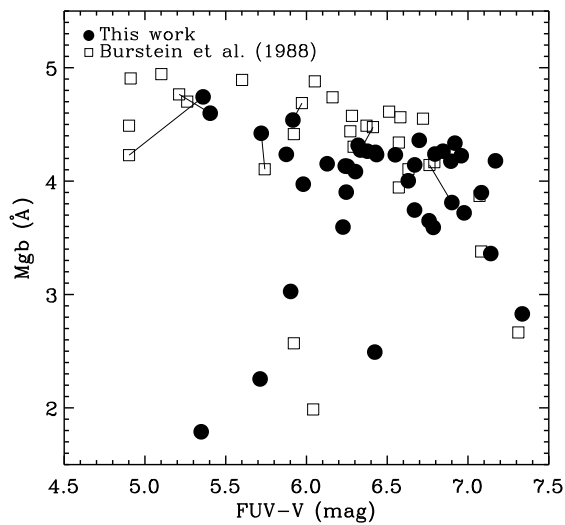}
\caption{(FUV$-V$)--\mgb\ (i.e.\ Burstein) relation for both our data
  (filled circles) and the 1550~\AA\ and Mg$_2$ data of
  \citeauthor{bbbfl88} (open squares). The latter have been transformed
  to FUV and \mgb\ as discussed in the text. Galaxies in common
  between the two samples are connected by a solid line. Error bars
  have been omitted for clarity. The two datasets are consistent, with
  similar scatter.}
\label{fig:burstein}
\end{center}
\end{figure}

Given the above, is our (FUV$-V$)--\mgb\ relation merely similar to
the original Burstein relation or is it identical? In
Figure~\ref{fig:burstein}, we have strived to transform the data of
\citeauthor{bbbfl88} to our own parameters. First, using the {\it IUE}
spectra of \citeauthor{bbbfl88}, we calculated the 1550$-$FUV colour
of NGC~221 and NGC~4649. To calculate the \galex\ FUV magnitude of
every galaxy in \citeauthor{bbbfl88}'s sample, we then simply assumed
that each galaxy had one of two spectral shapes: either a normal
early-type galaxy spectrum like NGC~221, or a UV upturn spectrum like
that of NGC~4649. The \galex\ FUV magnitudes were then simply obtained
by FUV=1550$-$(1550$-$FUV)$_{\rm NGC221,4649}$, using
(1550$-$FUV)$_{\rm NGC221}$ for galaxies with (1550$-V$)$\,\ge3.4$
(normal galaxies) and (1550$-$FUV)$_{\rm NGC4649}$ for galaxies with
(1550$-V$)$\,<3.4$ (UV upturn galaxies). As the {\it IUE} apertures of
\citeauthor{bbbfl88} are not that different from our own, we do not
correct the photometry for aperture mismatch. Second, we converted
\citeauthor{bbbfl88}'s Mg$_2$ index measurements into \mgb\ using the
\citet*{tmb03} stellar population synthesis models, assuming an
$\alpha$-element (over-)abundance [$\alpha$/Fe]=0.3 for all
galaxies. As those Mg$_2$ values are central, however, we also need to
correct the linestrengths for aperture mismatch. To do this, we simply
compared the transformed \citeauthor{bbbfl88} \mgb\ values with ours
for the $6$ galaxies in common (NGC~2768, NGC~4278, NGC~4374,
NGC~4486, NGC~4552, NGC~5846), and then applied an average offset to
all the galaxies ($\Delta$\mgb$=0.52$~\AA).
%
%

The conversion of \citeauthor{bbbfl88}'s data to our own parameters is
necessarily very approximate, but Figure~\ref{fig:burstein} shows that
our results are consistent with those of \citeauthor{bbbfl88}. Using
much improved data and measuring methods, we have thus recovered the
original Burstein relation. Somewhat disappointingly, however, the
scatter of our data is about the same. The scatter is also larger than
our measurement errors, which suggests that it is real. We explore
possible reasons why other studies did not recover the Burstein
relation in Sections~\ref{sec:sf} and \ref{sec:sdss}.
%
%
\subsection{FUV$-V$ vs.\ Fe5015}
\label{sec:corrs_fuv-v-Fe}
The middle panel of Figure~\ref{fig:fuv-v} shows the relation of
FUV$-V$ with Fe5015. The scatter is very large and a correlation
analogous to that of FUV$-V$ vs.\ \mgb\ is not obviously present.
This is perhaps not surprising, since the correlation of \mgb\ with
the stellar velocity dispersion $\sigma$ is also much stronger than
that of Fe5015. Equally unsurprising then, to our knowledge this
correlation has not been discussed much if at all in the literature.
Interestingly, however, galaxies with a young population now appear
both below and slightly above the best fit to the purely old objects
(especially for the smaller $R_{\rm e}/8$ apertures that are not
shown here).
%
%
\subsection{FUV$-V$ vs.\ \hb}
\label{sec:corrs_fuv-v-hbeta}
The bottom panel of Figure~\ref{fig:fuv-v} shows the relation of
FUV$-V$ with \hb, which is primarily influenced by age. Separating
young and old galaxies in this diagram is somewhat artificial, since
we use \hb\ itself as our age tracer, but it is striking that a tight
correlation is again only present amongst the galaxies with an old
stellar population.

Having said that, the observed variation of \hb\ with FUV$-V$ is
expected from the metallicity dependence of the latter (as traced
e.g.\ by the (FUV$-V$)--\mgb\ correlation). It thus does not
automatically imply an age dependence.
%
%
\subsection{NUV$-V$}
\label{sec:corrs_nuv-v}
\begin{figure}
\begin{center}
\includegraphics[width=7.45cm,clip=]{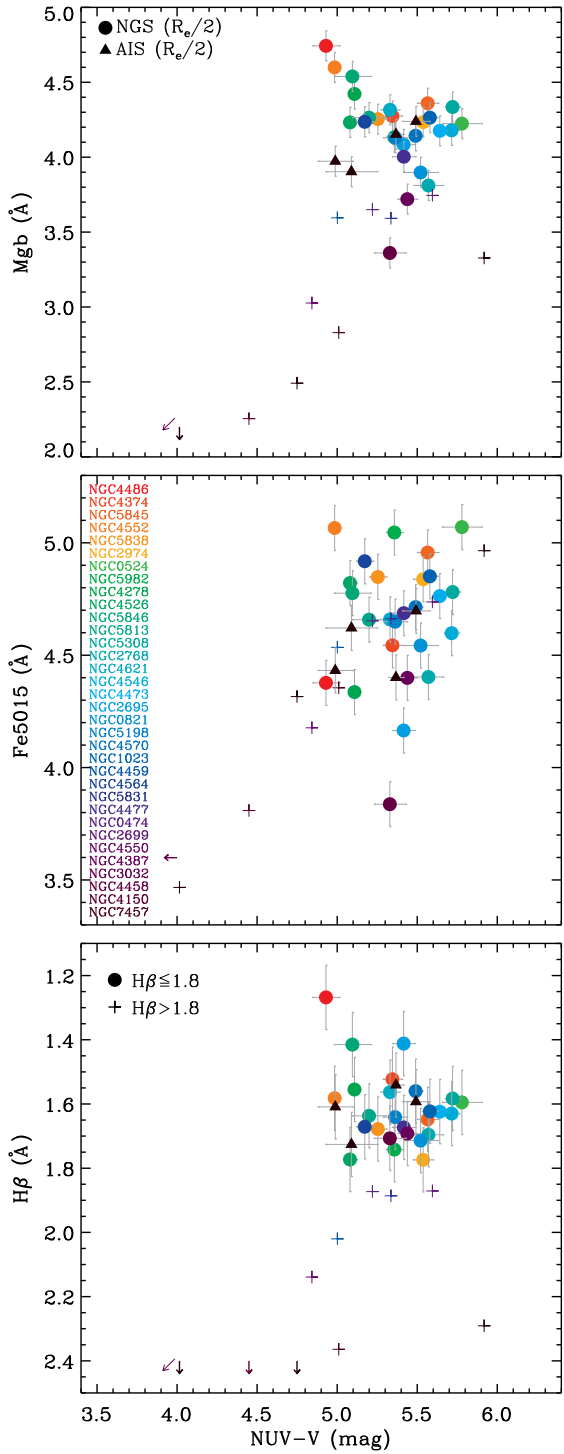}
\caption{Same as Figure~\ref{fig:fuv-v} but for the optical--UV colour
  (NUV$-V$).}
\label{fig:nuv-v}
\end{center}
\end{figure}

Similarly to Figure~\ref{fig:fuv-v}, Figure~\ref{fig:nuv-v} shows the
correlations between the NUV$-V$ colour and the \mgb, Fe5015 and \hb\
linestrength indices. The contrast between Figure~\ref{fig:fuv-v} and
\ref{fig:nuv-v} is however striking. First, the NUV$-V$ colour range
is only about half that in FUV$-V$ for old galaxies. Second, even when
considering only old galaxies, the UV--linestrength correlations have
essentially disappeared. Third, the galaxies with a young stellar
component are much more discrepant.
We also point out that nearly all galaxies with very blue NUV$-V$
colours also have high \hb\ linestrengths (as expected from star
formation), although the opposite is not always true.
%
%

Although present across the NUV filter, the UV upturn phenomenon
usually clearly dominates the spectral energy distribution only at
wavelengths blueward of $2000$~\AA. It thus cannot by itself entirely
explain the NUV properties observed. As suggested for example by
\citeauthor{detal07} (\citeyear{detal07}; but see also
\citealt{dor95}), who explored slightly different NUV$-V$
correlations, we believe that the above trends result from two main
competing effects. On the one hand, young stars have a greater effect
in the NUV than the FUV, as only very young stars ($t\la0.1$~Gyr) can
make contributions to the FUV light, while young stars in a much large
age range ($t\la1$~Gyr) contribute to the NUV. Young stars can also
influence \hb\ for even longer periods ($t\ga1$~Gyr). As testified by
the (NUV$-V$)--\hb\ plot, this is without doubt the reason why the
galaxies with a young stellar population are so discrepant. On the
other hand, line blanketing (the suppression of the continuum due to a
large number of partially overlapping absorption lines) is substantial
in the NUV and increases with metallicity. This is probably why the
FUV correlations largely disappear in the NUV even for galaxies with
no sign of significant recent star formation, as early-type galaxies
generally have a high metallicity. Those combined effects probably
also explain the lack of correlations in studies which lack a true FUV
filter \citep[e.g.][]{dbd02}.
%
%
\subsection{FUV$-$NUV}
\label{sec:corrs_fuv-nuv}
\begin{figure}
\begin{center}
\includegraphics[width=7.45cm,clip=]{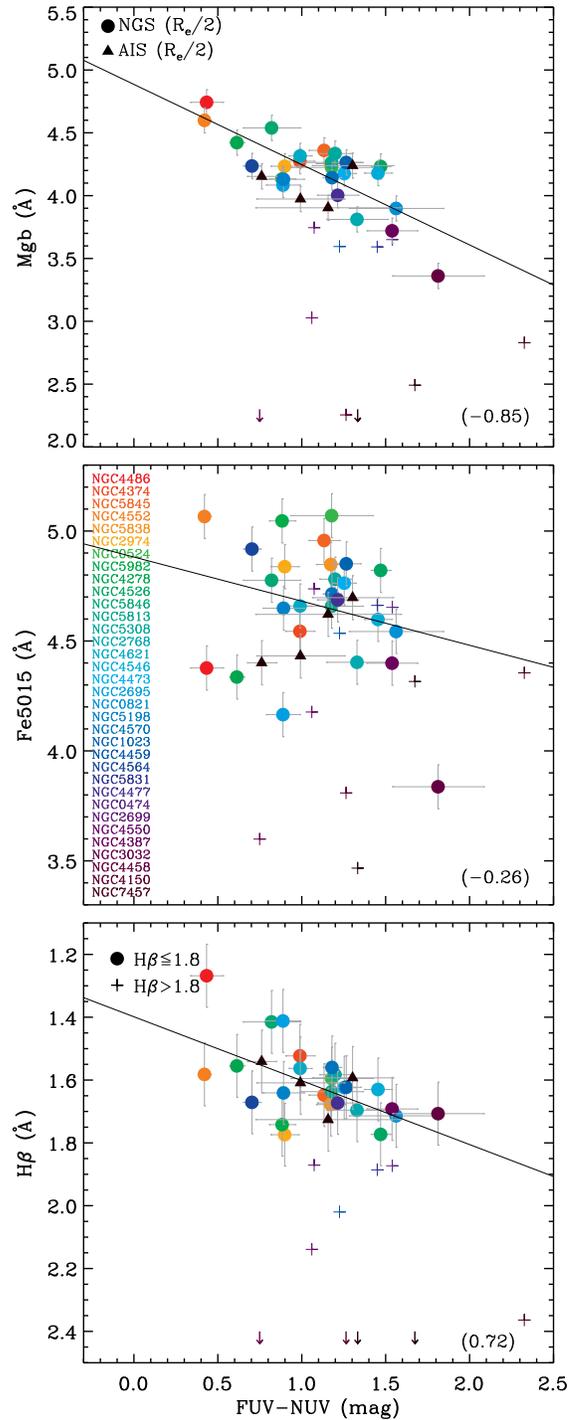}
\caption{Same as Figure~\ref{fig:fuv-v} but for the pure UV colour
  (FUV$-$NUV).}
\label{fig:fuv-nuv}
\end{center}
\end{figure}

Analogous to Figures~\ref{fig:fuv-v} and \ref{fig:nuv-v},
Figure~\ref{fig:fuv-nuv} shows the correlations between the FUV$-$NUV
colour and the \mgb, Fe5015 and \hb\ linestrength indices.
Interestingly, as pointed out especially by \citet{detal07}, the
UV--linestrength correlations are actually much more pronouced and
show less scatter when the UV$-$optical colour FUV$-V$ is replaced by
the pure UV colour FUV$-$NUV. As the metallicity increases, this is
most likely due to an increase in the FUV upturn conspiring with
increasing line blanketing in the NUV. However, given that the NUV
atmospheric opacity is dominated by iron-peak elements, it is
surprising that the correlation with Fe5015 is poorer than that with
\mgb. As expected, however, the FUV$-$NUV colours of galaxies with a
significant young stellar population are largely uncorrelated with the
linestrengths. The parameters of the best-fit linear relations are
again listed in Table~\ref{tab:linear_fits}.

We stress here that the (FUV$-$NUV)--\mgb\ correlation is tighter than
the Burstein relation, which uses FUV$-V$. Although the Burstein
relation has traditionally been used in most work, presumably for
historical reasons dating back to \citeauthor{bbbfl88}, and it remains
important, our results suggest that (FUV$-$NUV)--\mgb\ should instead
be the correlation of choice. As FUV and NUV are generally measured
with the same instrument, using FUV$-$NUV rather than FUV$-V$ also
avoids potential uncertainties in the UV to optical flux calibration.
%
%

%
%
\begin{figure}
\begin{center}
\includegraphics[width=7.45cm,clip=]{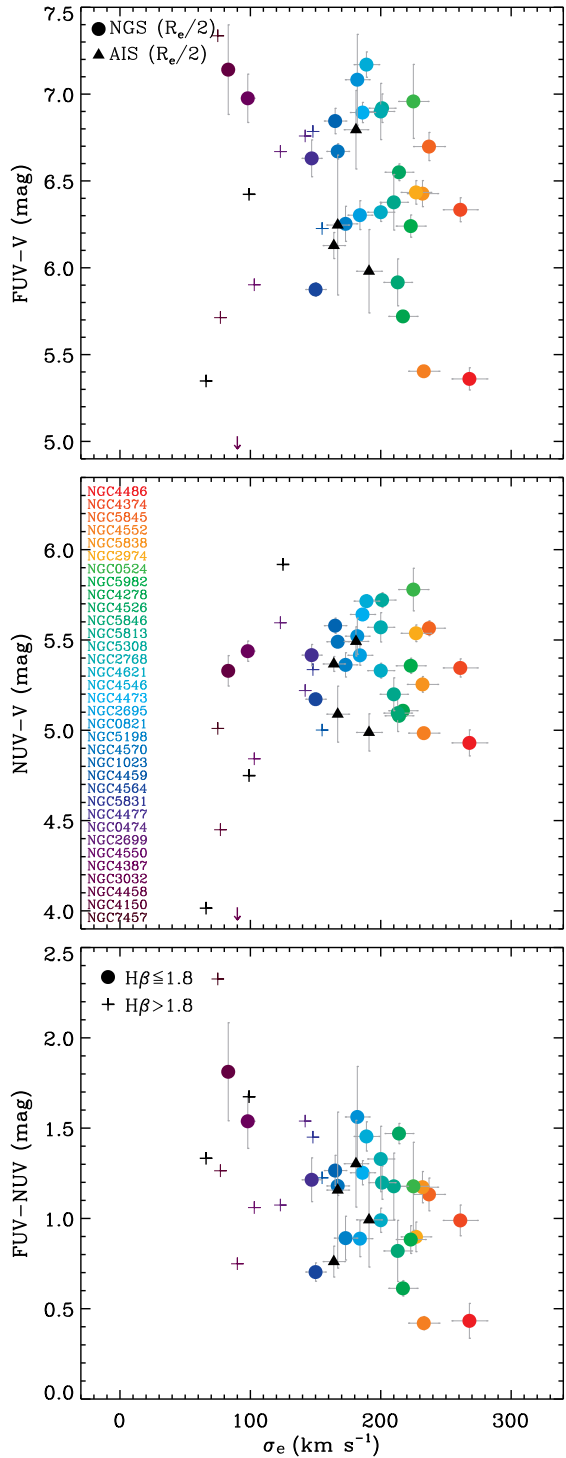}
\caption{Same as Figure~\ref{fig:fuv-v} but for the FUV$-V$, NUV$-V$
  and FUV$-$NUV colours as a function of the effective stellar
  velocity dispersion $\sigma_{\rm e}$.}
\label{fig:uv-sigma}
\end{center}
\end{figure}
%
%
\subsection{Stellar velocity dispersion}
\label{sec:corrs_dispersion}
The Burstein relation and other correlations highlighted in this
paper, combined with the well-known Mg--$\sigma$ relation
\cite*[e.g.][]{tdfb81,bbf93,cbdmsw99,betal03}, imply a correlation of
the UV-upturn phenomenon with the stellar velocity dispersion, which
is a good measure of a galaxy's gravitational potential depth and
total mass. The colour scheme in Figures~\ref{fig:fuv-v},
\ref{fig:nuv-v} and \ref{fig:fuv-nuv} allows us to gauge this
dependence, but only very roughly. It is thus shown explicitly in
Figure~\ref{fig:uv-sigma}. The corresponding figure showing the
dependence of the UV-upturn on the dynamical mass $M_{\rm
  dyn}\equiv5.0\,G^{-1}\,R_{\rm e}\,\sigma^2_{\rm e}$ (see
\citeauthor{cetal06}) is essentially identical. This mass represents
$M_{\rm dyn}\approx2\,M_{1/2}$, where $M_{1/2}$ is the total mass
within a sphere containing half of the galaxy light.

Comparing Figures~\ref{fig:fuv-v} and \ref{fig:uv-sigma}, one can see
that the scatter in the Burstein relation is smaller than that in the
(FUV$-V$)--$\sigma_{\rm e}$ relation. This effect first pointed out by
\citeauthor{bbbfl88} lead them to suggest that the UV upturn is
primarily driven by stellar population properties rather than
structural or dynamical ones. We confirm this trend here. The effect
is however much more pronounced when considering the FUV$-$NUV colour,
where the correlation with \mgb\ is very tight but that with
$\sigma_{\rm e}$ is much looser. Just as for \mgb, NUV$-V$ does not
show any clear dependence on $\sigma_{\rm e}$.

As expected from cosmic downsizing
\citep[e.g.][]{cshc96,teld05,tmbm05}, all galaxies with evidence of
recent star formation (high \hb) are in the low mass half of our
sample ($\sigma_{\rm e}\la180$~km~s$^{-1}$; see also
\citeauthor{shetal10}).
%
%
\subsection{Aperture effects}
\label{sec:corrs_aperture}
Analogous UV--linestrength correlations to those discussed in this
section are also present when using matching elliptical apertures of
$R_{\rm e}/8$ and $R_{\rm e}$. The linestrengths and broadband colours
are simply slightly offset, as expected from the radial stellar
populations gradients present in early-type galaxies. However, some
correlations are clearer for smaller apertures. This is most likely
because stellar age gradients are generally most pronounced in galaxy
centres (e.g.\ \citeauthor{ketal06}; \citealt*{metal06}, hereafter
\citeauthor{metal06}; \citeauthor{ketal10}). The integrated UV$-V$
colours and linestrengths thus have much broader ranges of values for
small apertures.

%
%
\begin{figure*}
\begin{center}
\includegraphics[width=0.7\textwidth,clip=]{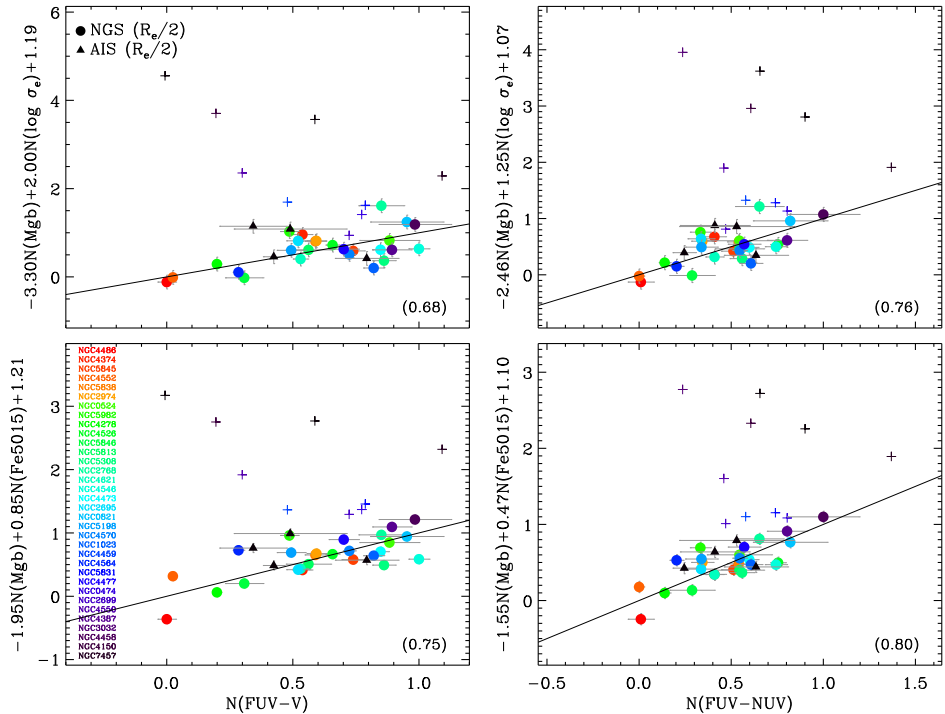}
\caption{Edge-on views of the best-fit planes for selected
  three-prameter relations (see text). {\em Top-left:} Plane relating
  FUV$-V$, \mgb\ and $\sigma_{\rm e}$. {\em Bottom-left:} Plane
  relating FUV$-V$, \mgb\ and Fe5015. {\em Top-right:} Plane relating
  FUV$-$NUV, \mgb\ and $\sigma_{\rm e}$. {\em Bottom-right:} Plane
  relating FUV$-$NUV, \mgb\ and Fe5015. The solid lines are shown to
  guide the eye and all have slope unity and go through the
  origin. The correlation coefficient is reported in the bottom-right
  corner of each panel. The three-parameter correlations shown are not
  tighter than the two-parameter correlations shown in
  Figures~\ref{fig:fuv-v} and \ref{fig:fuv-nuv}.}
\label{fig:multi}
\end{center}
\end{figure*}
%
%
\subsection{Multi-dimensional relations}
\label{sec:multi}
The various correlations shown in
Figures~\ref{fig:fuv-v}--\ref{fig:uv-sigma} suggest that \mgb\ is the
dominant parameter affecting the UV flux, but we now consider
relations involving three parameters, analogously to the Fundamental
Plane for stellar dynamics (relating the stellar luminosity, surface
brightness and central velocity dispersion of galaxies; see
\citealt{dd87,dlbdftw87}). Without an a priori physical foundation, such
as the virial theorem for the Fundamental Plane, it is however not
clear what correlations one should expect. To minimise biases arising
from the very different definitions and thus dynamic ranges of the
parameters considered (e.g.\ colours, velocity dispersion and
linestrengths), we normalize each parameter to yield an observed range
of $1.0$. For example, the measured values of FUV$-V$ extend from
roughly $5.4$ to $7.2$~mag (yielding an observed range of $1.8$~mag),
so in the fits below N(FUV$-V$)$=0.0$~mag is actually $1.8$~mag bluer
(UV stronger) than N(FUV$-V$)$=1.0$~mag, where the prefix N denotes a
normalised quantity. Likewise for the other quantities
(FUV$-$NUV$=0.4$--$1.8$~mag, $\log(\sigma_{\rm
  e}/$km~s$^{-1})=1.9$--$2.4$, \mgb$\,=3.4$--$4.7$~\AA,
Fe5015$\,=3.8$--$5.1$~\AA).

We have explored a number of parameter combinations, but we present
only the most promising ones below and in Figure~\ref{fig:multi}, in
the sense that the resulting planes (in their respective
three-dimensional spaces) are the tightest. The planes were obtained
by minimising the square of the residuals in the direction
perpendicular the planes, but we warn that the coefficients obtained
do depend sensitively on the exact quantity that is minimised. The
edge-on projections of the planes are shown in
Figure~\ref{fig:multi}. By construction, the solid lines shown (of
slope unity and going through the origin) should go through the data
points (they do). As before, the fits are performed to the quiescent
galaxies only (i.e.\ \hb$\,\le1.8$~\AA), yielding
\begin{equation}
{\rm N}({\rm FUV}-V)=-3.30\,{\rm N}({\rm Mg}\,b)+2.00\,{\rm N}(\log\sigma_{\rm e})+1.19\,\,,
\end{equation}
\begin{equation}
{\rm N}({\rm FUV}-V)=-1.95\,{\rm N}({\rm Mg}\,b)+0.85\,{\rm N}({\rm Fe5015})+1.21\,\,,
\end{equation}
\begin{equation}
{\rm N}({\rm FUV}-{\rm NUV})=-2.46\,{\rm N}({\rm Mg}\,b)+1.25\,{\rm N}(\log\sigma_{\rm e})+1.07\,\,,
\end{equation}
\begin{equation}
{\rm N}({\rm FUV}-{\rm NUV})=-1.55\,{\rm N}({\rm Mg}\,b)+0.47\,{\rm N}({\rm Fe5015})+1.10\,\,.
\end{equation}

The correlation coefficients reported in the bottom right corner of
each panel of Figure~\ref{fig:multi} are no better than those of
the good two-parameter fits shown in Figures~\ref{fig:fuv-v} and
\ref{fig:fuv-nuv}. None is better than the (FUV$-$NUV)--\mgb\
correlation. This implies that the FUV excess is not particularly
better described by employing an additional parameter. This is in
agreement with our finding of \S~\ref{sec:corrs_dispersion}, that it
is stellar population rather than dynamical properties that govern the
UV strength.
%
%

\begin{figure*}
\begin{center}
\includegraphics[width=0.7\textwidth,clip=]{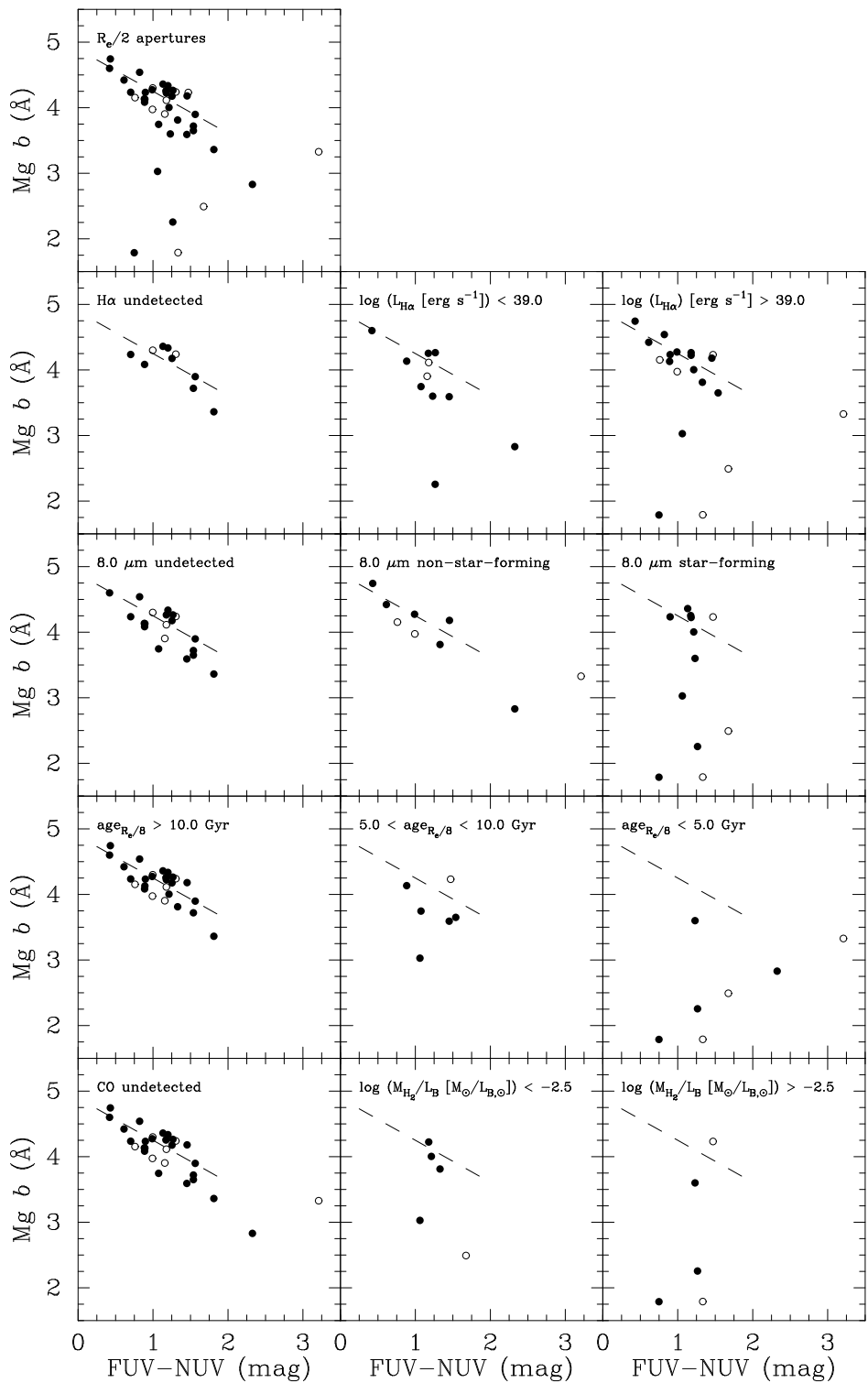}
\caption{Dependence of the (FUV$-$NUV)--\mgb\ correlation on recent
  and ongoing star formation. The top panel shows the correlation for
  all galaxies using elliptical apertures of $R_{\rm e}/2$. Lower
  panels show the correlation as a function of the integrated
  H$\alpha$ luminosity, the type of $8.0$~$\mu$m non-stellar emission
  if present, the luminosity-weighted mean stellar age and the
  molecular gas mass normalised by the $B$-band luminosity. From
  left-to-right, the columns respectively show galaxies with no, mild
  or ambiguous, and strong evidence for ongoing or recent star
  formation. A clear correlation with no outlier is only observed for
  quiescent galaxies. Filled symbols indicate galaxies with at least
  one orbit of \galex\ integration time, while open symbols indicate
  the shorter exposure AIS data. Error bars have been omitted for
  clarity and our best-fit (FUV$-$NUV)--\mgb\ relation is overplotted
  as a dashed line for reference.}
\label{fig:sf}
\end{center}
\end{figure*}
%
%
\section{STAR FORMATION}
\label{sec:sf}
In the (FUV$-V$)--linestrength and (FUV$-$NUV)--linestrength relations
discussed in Section~\ref{sec:corrs}, the apparent correlations
defined by the bulk of the observations are weakened by the presence
of a number of outlying data points. However, the outliers
systematically have \hb$\,>1.8$~\AA, which suggests that the
deviations from the bulk are due to the presence of a young stellar
component in these galaxies. Indeed, while often negligible in the
optical, even a small amount of recent star formation can have a
dramatic effect in the UV. We test this hypothesis thoroughly in this
section.

Figure~\ref{fig:sf} shows how the (FUV$-$NUV)--\mgb\ correlation
measured within $R_{\rm e}/2$ elliptical apertures depends on a number
of star formation tracers, none of which is however perfect. Our
best-fit (FUV$-$NUV)--\mgb\ relation from
Section~\ref{sec:corrs_fuv-nuv} is overplotted as a dashed line for
reference. The H$\alpha$ emission line and the mid-infrared (MIR)
emission from dust and polycyclic aromatic hydrocarbons (PAHs) are
normally considered good tracers of respectively unobscured and
obscured ongoing star formation, while the luminosity-weighted mean
stellar age traces recent and older star formation events. The
molecular gas mass $M_{{\rm H}_2}$ measures the fuel reserve for
current and future star formation. As we must rely on external sources
for those quantities, they are not measured within our standard
$R_{\rm e}/2$ aperture, but this is unimportant for the argument. We
only use those quantities to gauge the likelyhood of young stars; the
data plotted remain measured within identical apertures. The H$\alpha$
luminosity is integrated within the \sauron\ FOV and comes from
\citeauthor{setal06}. In the MIR, we use the {\it Spitzer}/Infrared
Array Camera (IRAC) $8.0$~$\mu$m non-stellar emission calculated and
discussed in \citeauthor{shetal10}. Considering the emission
morphology, the radial gradient of the $8.0$-to-$3.6$~$\mu$m flux
ratio, and the {\it Spitzer}/Infrared Spectrograph (IRS) spectrum when
available, \citeauthor{shetal10} very carefully separated the
$8.0$~$\mu$m non-stellar emission originating from star formation and
other processes. The mean stellar age is derived through single
stellar population synthesis models from linestrengths integrated
within a circular aperture of $R_{\rm e}/8$
(\citeauthor{ketal10}). The distance-independent $M_{{\rm H}_2}/L_B$
ratios are global values and come from the CO measurements of
\citet{cyb07} and \citet{yetal10} combined with the RC3 optical
catalogue \citep{rc3}.

The main conclusion from Figure~\ref{fig:sf} is that a clear, tight
(FUV$-$NUV)--\mgb\ correlation without outliers is only present for
galaxies which show no sign of current or recent star formation. The
scatter is much increased for galaxies with even mild evidence of star
formation, and essentially no correlation is present for galaxies with
strong evidence. This is particularly nicely illustrated by the mean
luminosity-weighted stellar age panels, which show increasing scatter
with decreasing mean age.

Other star formation tracers not shown in Figure~\ref{fig:sf} display
analogous behaviours, such as the \hb\ absorption linestrength and
NUV$-V$ colour discussed in this paper, the $B-V$ colour, and the
$8.0$~$\mu$m non-stellar luminosity and associated star formation
rate. More importantly, similar results are obtained for other
apertures ($R_{\rm e}/8$ and $R_{\rm e}$), FUV$-V$ and NUV$-V$
colours, and Fe5015 and \hb\ linestrengths. In addition, the
(FUV$-V$)--(FUV$-$NUV) and (FUV$-V$)--(NUV$-V$) colour-colour
relations clearly show that, while the quiescent galaxies
(\hb$\,\le1.8$~\AA) do define clear correlations \cite[see,
e.g.,][]{betal05,detal07}, galaxies with evidence of young stars
(\hb$\,>1.8$~\AA) systematically depart from them. This shows beyond
doubt that most deviations from the correlations described in
Section~\ref{sec:corrs} are due to galaxies with either current or
recent star formation, and that the correlations themselves are
essentially due to quiescent galaxies (i.e.\ purely old stellar
populations).

One should not be overly concerned that, occasionally, galaxies with
apparent signs of current or recent star formation in
Figure~\ref{fig:sf} do fall within the region expected of our best-fit
(FUV$-$NUV)--\mgb\ relation. As mentioned above, none of the tracers
is perfect, especially in early-type galaxies. In particular, as shown
in \citeauthor{setal10}, emission from Balmer lines such as H$\alpha$
in early-type galaxies is normallly dominated by non-star-formation
processes, specifically hot evolved stellar populations (PAGB
stars). Absence of H$\alpha$ emission thus guarantees absence of star
formation, but the presence of H$\alpha$ emission does not imply the
presence of star formation. Dust and PAHs can also be heated/ionised
by a variety of processes, including these same evolved stellar
populations (see \citeauthor{shetal10}). A (FUV$-$NUV)--\mgb\
correlation is only present for galaxies with $8.0$~$\mu$m non-stellar
emission if that emission has a non-star formation origin.

We also note that the scatter in our correlations is not caused by the
presence of active galactic nuclei (AGN). First, our measurements are
global rather than nuclear. Second, there are only four strong radio
continuum sources in the \sauron\ sample: two FR~I sources (NGC~4374
and NGC~4486), NGC~4278 and NGC~4552. None is particularly discrepant
in our correlations. It is clear however that AGN could contribute to
the scatter in samples of galaxies constructed differently.
%
%
\section{SDSS SAMPLE}
\label{sec:sdss}
\subsection{Sample selection and data}
\label{sec:sdss_sample}
We now place our findings based on observations of \sauron\ early-type
galaxies in the context of previous work on SDSS galaxies, such as
that of \citet{retal05}. UV and linestrength measurements of SDSS
early-type galaxies present a number of challenges. First, most
galaxies are invariably unresolved by \galex\ but the spectral
information is derived from SDSS fibre spectra with a $3\arcsec$
diameter, leading to aperture mismatch. Second, except at the highest
luminosities, many early-type galaxies in the low-redshift regime
($z\approx0.05$--$0.1$) are either undetected or barely detected at
the depth of the \galex\ Medium Imaging Survey (MIS;
\citealt{maetal05,moetal05}). Third, a reliable morphological
classification is far more challenging. Fourth, disentangling the UV
upturn from residual star formation is very difficult.

Given that we do find a correlation between UV$-$optical colours and
\mgb\ in \sauron\ galaxies, we nevertheless attempt here to overcome
these hurdles and show that the Burstein relation pioneered by
\citeauthor{bbbfl88} can also be found in more distant early-type
galaxies, as can other UV--linestrength relations. We use for this
purpose the Morphologically Selected Ellipticals in SDSS sample
(MOSES; \citealt{stsmkjys07}), which includes all spectroscopic SDSS
galaxies with $r<16.8$ in the redshift range $0.05<z< 0.1$ and
provides reliable visual morphologies.

The SDSS spectra were first processed with the software {\small
  GANDALF} to remove nebular emission lines (see
\citeauthor{setal06}), while using pPXF to fit the stellar kinematics
\citep{cc04}. The cleaned spectra were then degraded to the spectral
resolution of the Lick indices and stellar absorption linestrengths
were measured, including \mgb\ and \hb. For details, see
\citet{stsmkjys07} and \citet{tmssjkys09}. With those data in hand, we
now address the four challenges listed above.

First, there is no reasonable correction we can apply to the issue of
aperture mismatch, so this necessarily remains a source of systematic
error for \galex-SDSS galaxies.

Second, since the slope of the Burstein relation is not particularly
steep, there is a danger that it could be washed out by large errors
in the UV photometry. Even though these requirements substantially
reduce the sample size, we thus only consider here those MOSES
early-type galaxies that are detected by \galex\ in {\it both} NUV and
FUV filters, and we further require that the photometric error in each
be less than $0.3$~mag. These two requirements reduce the sample from
$12\,828$ galaxies for MOSES to $535$. Perhaps more importantly, these
requirements strongly bias our sample in favour of UV-blue galaxies,
particularly compared to the \sauron\ galaxies discussed in this
paper. UV-red galaxies are largely excluded.

Third, it is clear that any automated early-type galaxy selection
criteria based on structural parameters leads to the inclusion of many
disc-dominated galaxies, Sa galaxies being a particularly prevalent
source of contamination \cite[see, e.g.,][]{setal07}. As galaxies with
discs are likely forming stars, they will contaminate the Burstein and
other similar relations with UV flux from young stars. The most
reliable method for excluding such non-early-type interlopers is
visual inspection, which has been carefully performed for the MOSES
sample using SDSS images. Short of obtaining space-based data for a
substantial number of these galaxies, this is thus the best
morphological classification possible.

Fourth, even after removing intermediate and late-type interlopers,
residual star formation is still common in low-redshift early-type
galaxies \citep[e.g.][]{yetal05,setal07,ketal07}, and this residual
star formation will dilute the Burstein and other relations. To remove
these objects from the sample, it is possible to reject all galaxies
with high \hb\ linestrengths. Specifically, we investigate here
rejecting objects with \hb\ $+$ \hb$_{\rm err}>1.8$~\AA\ (see below).
\subsection{Burstein relation: FUV$-V$ vs.\ \mgb}
\label{sec:sdss_burstein}
\begin{figure}
\begin{center}
\includegraphics[height=0.45\textwidth,angle=90.0,clip=]{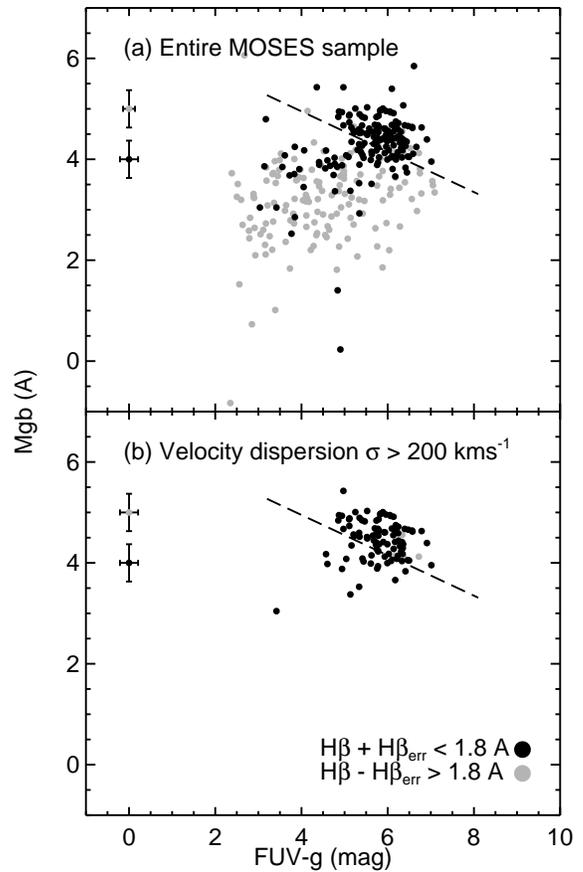}\\
\caption{Mg--(FUV$-g$) relation for the MOSES sample early-type
  galaxies with good UV photometry. {\em Top:} Entire sample meeting
  the \hb\ selection criteria. {\em Bottom:} Only galaxies with a
  velocity dispersion $\sigma$ larger than
  $200$~km~s$^{-1}$. Quiescent galaxies with \hb+\hb$_{\rm
    err}\le1.8$~\AA\ are shown as black circles, while galaxies
  consistent with harbouring a young stellar population (\hb-\hb$_{\rm
    err}>1.8$~\AA) are shown in grey. Characteristic error bars are
  shown in the top-left of each plot. Our best-fit Burstein relation
  is overplotted as a dashed line for reference, taking into acount
  the small $g-V$ colour transformation described in the text.}
\label{fig:sdss}
\end{center}
\end{figure}

We plot in the top panel of Figure~\ref{fig:sdss} the \mgb\
linestrength against FUV$-g$ colour for the SDSS early-type galaxies
selected. Overplotted is the best-fit Burstein relation derived in
Section~\ref{sec:corrs_fuv-v-mgb}, transformed from $V$ to $g$. The
latter correction was obtained using a large sample of early-type
galaxies with \galex\ data selected visually from SDSS. As the $g-V$
colour effectively does not vary with FUV$-$NUV, we adopted a fixed
correction $g-V=0.53$~mag.
%
%

Ignoring the cut on \hb\ linestrength, the data in the top panel of
Figure~\ref{fig:sdss} only show a loose correlation in a sense
opposite to the Burstein relation discussed in
Section~\ref{sec:corrs_fuv-v-mgb}. The immense scatter introduced by
residual star formation largely washes out the signal in the
UV$-$optical colour. If we however separate the sample according to
\hb\ linestrength, the picture changes dramatically. Quiescent
early-type galaxies with low \hb\ values ($161$ objects with
\hb+\hb$_{\rm err}\le1.8$~\AA; black circles) cluster at redder
colours and are located roughly within the region expected from the
original Burstein relation, with few outliers. Galaxies with a young
population with high \hb\ values ($144$ objects with \hb-\hb$_{\rm
  err}>1.8$~\AA; grey circles) do not and have an entirely different
distribution. This selection on \hb\ does exclude galaxies whose error
bars bestride the \hb$\,=1.8$~\AA\ threshold between recently
star-forming and non-recently star-forming galaxies (i.e.\ galaxies
for which \hb-\hb$_{\rm err}\le1.8$~\AA\ $<\,$\hb+\hb$_{\rm err}$),
but we believe this is best suited to test our hypothesis that star
formation (i.e.\ the presence of young stars) is mainly responsible
for the outliers in the Burstein relation. Galaxies that bridge the
threshold may or may not contain young stars and will blur any
dichotomy.

Our \hb\ criteria are however not perfect discriminants between
entirely passive early-type galaxies and those with a low level of
residual star formation in the last gigayear. In any case, residual
star formation can occur at large radii (see \citeauthor{jetal09}),
and this will not be picked up by the SDSS fibres. If we use a blunter
selection criterion and restrict ourselves further to early-type
galaxies with a high velocity dispersion ($\sigma>200$~kms$^{-1}$;
$169$ of $535$ galaxies), where residual star formation is virtually
completely suppressed \citep[see, e.g.,]{scetal06}, then we can see
that {\em all} early-type galaxies cluster within the same locus
regardless of the \hb\ criteria (see Fig.~\ref{fig:sdss}, bottom
panel). And indeed, there are very few galaxies with \hb-\hb$_{\rm
  err}>1.8$~\AA\ left ($5$ objects compared to $97$ for \hb+\hb$_{\rm
  err}\le1.8$~\AA). The large errors in both \mgb\ and FUV$-g$, in
addition to aperture mismatch, result in sufficient scatter to largely
hide the (FUV$-g$)--\mgb\ relation, but the locus is consistent with
the slope and zero-point found for the \sauron\ galaxies.

Our treatment of the SDSS sample thus supports the presence of
(UV$-$optical)--linestrength relations as described in
Section~\ref{sec:corrs_fuv-v-mgb}, but again only for quiescent
galaxies. This is true using both an \hb\ linestrength (i.e.\ age) cut
and a stellar velocity dispersion (i.e.\ total mass) cut. Our
treatment also illustrates how challenging weeding out galaxies with
low levels of residual star formation can be with the data quality
typical of large surveys. Here, the sheer number of galaxies does not
trump the need for high quality data.
%
%
\section{DISCUSSION}
\label{sec:discussion}
\subsection{Origin of the UV upturn}
\label{sec:discussion_uvx}
A main result of our work illustrated in
Figures~\ref{fig:fuv-v}--\ref{fig:fuv-nuv} is that we recover the
Burstein relation for early-type galaxies originally discussed in
\citeauthor{bbbfl88}, as well as the (NUV$-V$)--\mgb\ and
(FUV$-$NUV)--\mgb\ correlations discussed in later papers
\citep[e.g.][]{dor95,retal05,betal05,detal07}. However, contrary to
these authors, we use identical apertures for all quantities. We have
also clearly showed that correlations with the \hb\ linestrength index
also exist, albeit usually with larger scatter. Correlations with
Fe5015 are at best very weak. These additional correlations are
crucial, as once compared to stellar population synthesis models they
will allow us to roughly separate the effects of $\alpha$-element
enhancement, metallicity and age on the UV-upturn phenomenon (see,
e.g., \citeauthor{ketal10}).

Lacking other elements, previous works considered Mg as a metallicity
tracer (e.g.\ \citeauthor{bbbfl88}; \citealt{retal05,detal07}). But
given that early-type galaxies and bulges are known to be
over-abundant in $\alpha$ elements \citep*[e.g.][]{r88,gea90,wfg92},
one might now naively be tempted to treat Mg primarily as a tracer of
$\alpha$-elements, and take Fe5015 as the prime metallicity tracer. A
high [$\alpha$/Fe] ratio is usually interpreted as implying a short
star formation timescale (i.e.\ a burst), as stars are enriched in
$\alpha$ elements produced by early Type~II supernovae but are not
enriched in Fe produced by longer timescale Type~Ia supernovae
\citep*[e.g.][]{eaglnt93,f98,tgb98}. Taken at face-value, the tight
(FUV$-V$)--\mgb\ correlation (i.e.\ the Burstein relation) would thus
suggest that the UV upturn in early-types is primarily due to stars
formed in a rapid burst and enriched in $\alpha$ elements. The
extremely loose (FUV$-V$)--Fe5015 correlation would further suggest
that enrichment in Fe produced over longer timescales is not the
primary driver of the correlations. This would then presumably pose
challenges to UV upturn theoretical models with strong metallicity
dependences (e.g.\ EHB models), and equally spell bad news for
scenarios relying on binary star evolution more generally
\citep[e.g.][]{mhmn01,hpmm03,hpl07}.

However, are we really justified to make this simplistic association
of \mgb\ with $\alpha$ elements and Fe5015 with metallicity? The short
answer is (un)fortunately no. Although Mg is an $\alpha$ element, the
\mgb\ index is still primarily driven by metallicity, while the Fe5015
index includes a number of poorly understood absorption
lines. Furthermore, while they slightly affect both, $\alpha$ elements
dominate neither the energy generation (He, CNO) nor the atmospheric
opacity (iron-peak elements) in metal-rich galaxies
\citep[e.g.][]{gzzh10}. Beside specific spectral features, their
effect on UV light should thus be limited \citep[see also][]{y08}. A
proper unraveling of the individual effects of $\alpha$-element
enhancement and metallicity on the UV upturn is thus carried out in
the companion paper by Jeong et al.\ (in preparation), with the help
of the stellar population synthesis models presented in
\citeauthor{ketal10}. This is not only the case for the primary
dependence of the UV upturn on metallicity, but also for its
dependence on \hb\ for old galaxies (\hb$\,\le1.8$~\AA). A priori,
this may well be explained entirely by metallicity effects, but it
could also hide a residual age dependence.

\begin{figure}
\begin{center}
\includegraphics[width=0.45\textwidth,clip=]{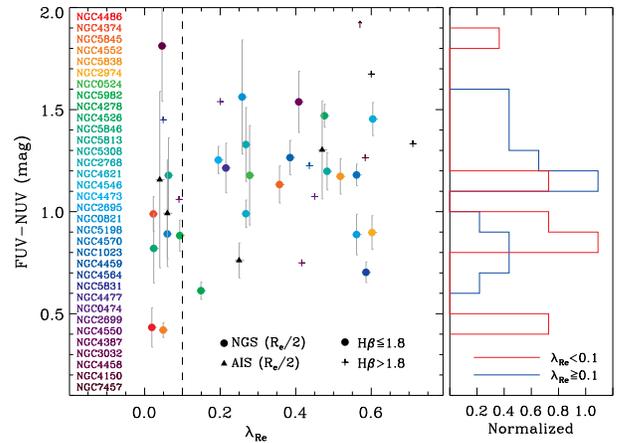}
\caption{Same as Figure~\ref{fig:fuv-v} but for the FUV$-$NUV colour
  as a function of the specific angular momentum of the galaxies
  $\lambda_{\rm R}$. The normalised number distribution of
  slow-rotator (red) and fast-rotator (blue) galaxies as a function of
  the FUV$-$NUV colour is shown on the right for quiescent galaxies only
  (i.e.\ cirles and triangles).}
\label{fig:lambda_R}
\end{center}
\end{figure}

Despite having failed to find a dominant dependence of the UV upturn
on $\sigma_{\rm e}$, we can test if other parameters might better
reveal a structural or dynamical influence on the UV upturn. We find
no or weak dependences of the FUV$-V$, NUV$-V$ and FUV$-$NUV colours
on the ellipticity $\epsilon$, boxiness of the isophotes $a_4$ and the
ratio of rotational to pressure support $V/\sigma$. In the spirit of
recent results of the \sauron\ survey, we plot in
Figure~\ref{fig:lambda_R} the dependence of the UV upturn on
$\lambda_{R_{\rm e}}$, the specific angular momentum of a galaxy
measured within $R_{\rm e}$ (see \citeauthor{eetal07} for
details). Figure~\ref{fig:lambda_R} also shows the normalised number
distribution of slow-rotator ($\lambda_{R_{\rm e}}<0.1$) and
fast-rotator ($\lambda_{R_{\rm e}}\ge0.1$) galaxies as a function of
the FUV$-$NUV colour, for quiescent galaxies only. Although there is
much scatter, we see that among quiescent galaxies very blue FUV$-$NUV
colours are more commonly observed among slow rotators than fast
rotators. Clearly, however, more extensive investigations are
necessary to substantiate any dynamical dependence of the UV upturn.
%
%

Lastly, we point out that the scatter in the UV--linestrength
relations presented here (e.g.\ the Burstein relation) is larger than
our measurement errors. It is thus real, implying that there is not a
one-to-one correlation between the UV excess and absorption
linestrengths. This is an important result, but stellar evolution
models of course do not depend on linestrengths but rather on age,
metallicity and $\alpha$-element abundance.
\subsection{Star formation}
\label{sec:discussion_sf}
Our work shows that star formation is responsible for most outliers in
the UV--linestrength correlations. Indeed, with the extensive
ancillary data available for the \sauron\ sample, and the exquisite
sensitivity reached for nearby objects, we were able to systematically
identify all outliers with galaxies showing signs of recent or current
star formation. This could also be done in an analogous manner for a
much larger (and more distant) sample of galaxies from SDSS. The fact
that clear correlations are then only recovered for massive galaxies
with high quality data can probably be explained by the relative poor
sensitivity of SDSS to the presence of recent and/or current star
formation (more likely present in low-mass systems).

The same reason probably explains the results of \citet{retal05},
which are most at odds with those of \citeauthor{bbbfl88}. We support
here the results of \citet{detal07}, who argued in favour of
contamination from low-level star formation in their own sample,
especially for the lenticulars. Requiring S/N$\ge10$ for the SDSS
spectra (although they quote a final S/N$=23\pm7$ for their quiescent
sample), \citet{retal05} were probably not as sensitive to weak
emission lines and thus star formation as we are here. It would be
interesting to see the UV$-$optical colour corresponding to their
H$\alpha$ emission (and thus star formation) upper limits. We require
here S/N$\ge60$ per spatial bin (and spectral element) for \sauron\
and simultaneously fit both absorption and emission lines, yielding an
equivalent width sensitivity of $0.1$~\AA\ (see
\citeauthor{setal06}). We also have much ancillary data, including the
age-sensitive \hb\ absorption line index.

We also stress here a fact that has generally been overlooked in
previous UV-upturn observational studies. All galaxies which are off
the relations, and which we identify as having current or recent star
formation in Sections~\ref{sec:sf} and \ref{sec:sdss}, are
systematically below the correlations. That is, they systematically
have \mgb\ and Fe5015 linestrengths which are lower than those
predicted by the correlations for purely old galaxies. None has
significantly higher \mgb\ or Fe5015, and the effect is particularly
pronounced for \mgb. Considering the predictions of stellar population
synthesis models \citep[e.g.][]{tmb03}, this is entirely consistent
with the outliers being of systematically lower mean ages.
%
%
\section{CONCLUSIONS}
\label{sec:conclusions}
We have used space-based UV ultraviolet photometry from \galex,
ground-based optical photometry from MDM and ground-based optical
integral-field spectroscopy from \sauron\ to study the
UV--linestrength relations of early-type galaxies. Focusing on the
\sauron\ sample, those data represent significant improvements over
previous work. In particular, identical apertures could be used for
all datasets, eliminating the aperture mismatch issues that have
affected previous studies.

We have shown that early-type galaxies follow fairly tight
correlations between the integrated FUV$-V$ and FUV$-$NUV colours and
the integrated \mgb\ and \hb\ linestrength indices, less so for the
NUV$-V$ colour, Fe5015 index and stellar velocity dispersion $\sigma$. Through stellar population
synthesis models, these correlations constrain the dependence of the
UV-upturn phenomenon in early-type galaxies on age, metallicity and
$\alpha$-element abundance. In particular, despite some recent
controversy stemming from the analysis of large survey datasets, we
have recovered a (FUV$-V$)--\mgb\ correlation that is analogous and
entirely consistent with that first proposed by
\citeauthor{bbbfl88}. We refer to this and the original relation as
the Burstein relation. This relation clearly suggests a positive
dependence of the UV-upturn on metallicity and is stronger when the
pure UV colour (FUV$-$NUV) is used, although any possible dependence
on age must await a better comparison with models. The scatter in the
correlations appears to be real and there is mild evidence that a
strong UV excess is preferentially present in slow-rotating galaxies,
although this last point deserves further study.

Using ancillary data constraining past and current star formation in
our sample galaxies, we have shown that most data that do not follow
the main correlations can be attributed to galaxies with evidence for
recent star formation, even if weak. The UV--linestrength correlations
discussed thus appear to hold exclusively for old stellar
populations. A large sample of more distant early-type galaxies
selected from SDSS, and thus easier to relate to current large
surveys, supports this view. The same can be said of the correlations
outliers themselves, which have systematically lower \mgb, Fe5015 and
higher \hb\ linestrength indices.

Given that galaxies show internal colour and linestrength gradients,
it is natural to ask whether the global correlations observed here
also apply locally within galaxies. Using the full capabilities
accorded to us by integral-field data, this issue is the focus of
a companion paper (Jeong et al., in preparation).
%
%
\section*{Acknowledgments}
{\bf [Dedication more exhaustive than that accepted by MNRAS]}\\
We dedicate this paper to the memory of David Burstein, who passed
away in December 2009. David was the leading author on the paper that
triggered our investigation, concerning the relationship between
magnesium linestrength and ultraviolet colours in galaxies, a subject
that remains enigmatic and the focus of active research today. During
his distinguised career, David made major contributions to several
fields of astronomy. His most cited work is on reddening estimates
determined from galaxy counts and H{\small I} column densities,
published with Carl Heiles in 1982.  That work was the primary source
of reddening estimates for over $15$ years. David's impact was rooted
in a deep understanding of the astronomical techniques in which he was
expert. His skill with photometry made him an invaluable member of the
`seven samurai' team that discovered large-scale motions of galaxies
and the `Great Attractor' in the 1980s. Equally, with colleagues at
the Carnegie Institution in Washington, David applied his expertise in
rotation curves to the understanding of spiral galaxies, work that was
pivotal in establishing the existence of dark matter haloes around
spirals. Throughout his career David worked on the subject of stellar
populations, making several groundbreaking advances as part of the
Lick team he first joined as a graduate student. In addition to being
a forefront researcher, David Burstein was a dedicated member of the
faculty at Arizona State University where he worked for $26$ years,
serving a term as President of the Academic Assembly. It is a
priviledge for us to have been able to contribute to the advancement
of knowledge in one area in which David worked, and which is now
commonly referred to as the `Burstein relation'. Ours will surely be
only one of many papers that build on David Burstein's scientific
legacy.

We thank the staff of the \galex\ project, MDM Observatory and Isaac
Newton Group for their assistance during and after the
observations. We would also like to thank S.\ Kaviraj and S.\ Trager
for useful discussions. The SAURON project is made possible through
grants from NWO and financial contributions from the Institut National
des Sciences de l'Univers, the Universit\'{e} Lyon I, the Universities
of Durham, Leiden and Oxford, the Programme National Galaxies, the
British Council, STFC grant 'Observational Astrophysics at Oxford' and
support from Christ Church Oxford and the Netherlands Research School
for Astronomy NOVA. MB acknowledges support from NASA through \galex\
Guest Investigator program GALEXGI04-0000-0109. SKY acknowledges
support to the Center for Galaxy Evolution Research from the National
Research Foundation of Korea, from a Korea Research Foundation Grant
(KRF-C00156) and from a Doyak grant (No.\ 20090078756). MB and RLD are
grateful for postdoctoral support through STFC rolling grant
PP/E001114/1. MB and SKY are also grateful to the Royal Society for an
International Joint Project award (2007/R2) supporting this work. The
STFC Visitors grant to Oxford also supported joint visits. KS was
supported by NASA through Einstein Postdoctoral Fellowship grant
number PF9-00069 issued by the Chandra X-ray Observatory Center, which
is operated by the Smithsonian Astrophysical Observatory for and on
behalf of NASA under contract NAS8-03060. KS also gratefully
acknowledges previous support from Yale University and a Henry Skynner
Junior Research Fellowship. MC acknowledges support from a STFC
Advanced Fellowship (PP/D005574/1). JFB acknowledges support from the
Ram\'{o}n y Cajal Program financed by the Spanish Ministry of Science
and Innovation. DK acknowledges support from Queen's College Oxford
and the hospitality of Centre for Astrophysics Research at the
University of Hertfordshire. Based on observations made with the NASA
Galaxy Evolution Explorer. \galex\ is operated for NASA by the
California Institute of Technology under NASA contract
NAS5-98034. Photometric data were also obtained using the 1.3m
McGraw-Hill Telescope of the MDM Observatory, and spectroscopic data
are based on observations obtained at the William Herschel Telescope,
operated by the Isaac Newton Group in the Spanish Observatorio del
Roque de los Muchachos of the Instituto de Astrofísica de
Canarias. Part of this work relied on data obtained from the
ESO/ST-ECF Science Archive Facility. We acknowledge the usage of the
HyperLeda database (http://leda.univ-lyon1.fr). This project also made
use of the NASA/IPAC Extragalactic Database (NED) which is operated by
the Jet Propulsion Laboratory, California Institute of Technology,
under contract with the National Aeronautics and Space Administration.
%
%

%
\appendix
%
%
\section{AVERAGE LINESTRENGTH MEASUREMENTS}
\label{app:averages}
Formally, to measure the average linestrength within a \sauron\
aperture containing multiple bins, we should sum the (fully
calibrated) spectra of all the bins within that aperture and
re-measure the linestrength. However, recalculating the linestrength
implies re-deriving the stellar kinematics, re-subtracting the ionised
gas emission, re-correcting the linestrength for velocity broadening,
etc, as done in \citeauthor{ketal06}. It is thus a time-consuming
task. Furthermore, in a companion paper exploring the UV--linestrength
relations as a function of radius within individual galaxies (Jeong et
al., in preparation), we integrate the linestrengths over elliptical
annuli (rather than elliptical apertures). The resulting LOSVDs can
then be significantly non-Gaussian, which makes the stellar kinematic
and velocity dispersion correction calculations problematic. It is
thus preferable to find a method to integrate (i.e.\ average) the
linestrengths in an aperture using the linestrength measurements in
individual bins themselves.

\begin{figure}
\begin{center}
\includegraphics[width=7.45cm,clip=]{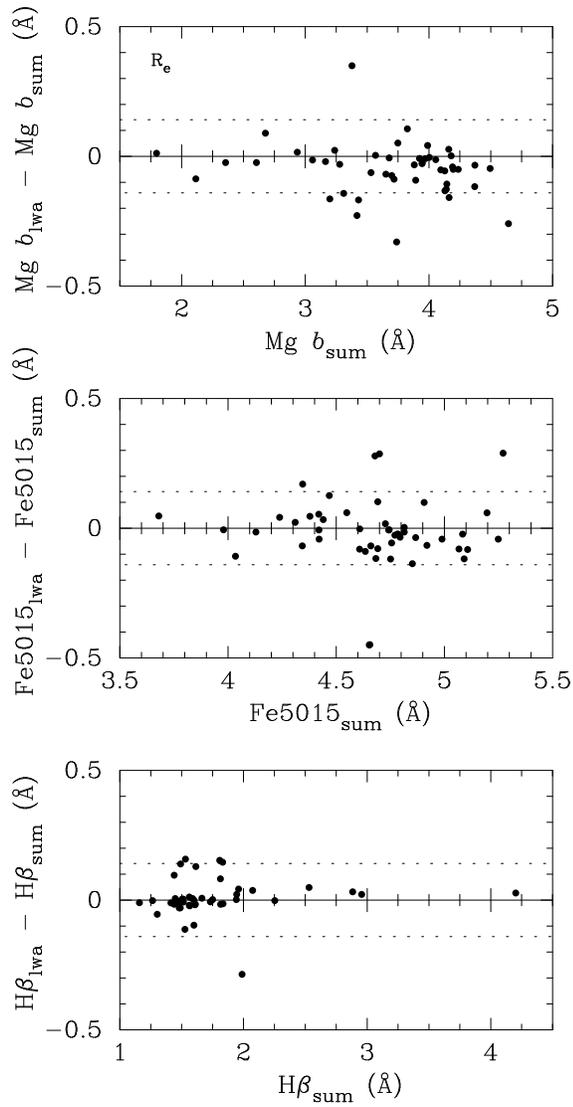}
\caption{Comparison of integrated linestrength values obtained by
  properly summing the spectra (labeled ``sum'') and simply
  luminosity-weighting the linestrength values themselves (labeled
  ``lwa'') within circular apertures of radii $R_{\rm e}$. Error bars
  are omitted for clarity, but the adopted formal random error on
  either quantity is $0.1$~\AA. {\em Top:} \mgb. {\em Middle:}
  Fe5015. {\em Bottom:} \hb. The scatter is always much smaller than
  the range of values covered and there is no significant systematic
  trend. The dotted lines show the $0.14$~\AA\ formal random error on
  the difference.}
\label{fig:average}
\end{center}
\end{figure}

In Figure~\ref{fig:average}, we show the properly averaged
linestrength measurements of \citeauthor{ketal10} within a circular
aperture of radius $R_{\rm e}$, compared to measurements obtained by
simply averaging the linestrength values of the individual bins within
that same aperture, weighted by the luminosity of each bin. While
there is some scatter, it is much smaller than the range of values
covered, the difference between the two measurements is always small,
typically less than the error on each measurement
(\citeauthor{ketal10} quotes a formal random uncertainty on integrated
linestrengths of at most $0.1$~\AA, and we have adopted the same
uncertainty on our measurements here), and the few outliers
systematically have very large effective radii. Most importantly,
there is no significant trend with the linestrength values
themselves. An aperture of $R_{\rm e}$ is also the worst case
scenario, and even smaller differences are obtained for smaller
apertures involving fewer bins. It thus seems that, although not
formally equivalent, simple luminosity-weighted linestrength averages
do yield accurate estimates. We have thus adopted this averaging
method for \sauron\ papers requiring integrated linestrength values
(e.g., \citealt*{setal09}, hereafter \citeauthor{setal09}; Jeong et
al., in preparation).
%
%
\end{document}